\documentclass{article}

\usepackage{arxiv}

\usepackage[utf8]{inputenc}
\usepackage[T1]{fontenc}
\usepackage[english]{babel}
\usepackage{microtype}

\usepackage{amsmath}
\usepackage{amsfonts}
\usepackage{amssymb}
\usepackage{nicefrac}
\usepackage{mathtools}

\usepackage{graphicx}
\usepackage{booktabs}
\usepackage{array}
\usepackage{tabularx}
\usepackage{xcolor}
\usepackage{ragged2e}

\usepackage{tikz}
\usetikzlibrary{
  arrows.meta, positioning, shapes.geometric, shadows, calc,
  decorations.pathreplacing, fit, backgrounds, shapes.misc
}

\usepackage[numbers,sort&compress]{natbib}
\usepackage{hyperref}
\usepackage{url}
\usepackage{cleveref}
\usepackage{enumitem}
\usepackage{authblk}
\usepackage{orcidlink}
\usepackage{placeins}
\usepackage{float}

\definecolor{rulecol}{rgb}{0.10,0.30,0.55}
\definecolor{accent1}{rgb}{0.85,0.33,0.10}
\definecolor{accent2}{rgb}{0.18,0.49,0.20}
\definecolor{neutral}{rgb}{0.40,0.40,0.40}

\graphicspath{{figures/}}

\title{Margin Play: A Multi-Agent System for Public Policy Analysis
  in the Brazilian Equatorial Margin}

\renewcommand{\shorttitle}{Margin Play: MARL for Public Policy Analysis in the BEM}

\author[1,3]{Antonio de Sousa Leitão Filho~\orcidlink{0009-0002-1705-3611}%
  \thanks{Corresponding author:
    \href{mailto:antonio@aiacontext.com}{\texttt{antonio@aiacontext.com}}}}
\author[1,3]{Fabrício Saul Lima}
\author[1]{Selby Mykael Lima dos Santos}
\author[1]{Rejani Bandeira Vieira Sousa}
\author[2]{Luís Jorge Mesquita de Jesus}
\author[3]{Dennys Correia da Silva}
\author[3]{Allan Kardec Duailibe Barros Filho}

\affil[1]{Aia Context, São Luís, MA, Brazil}
\affil[2]{Universidade Estadual de Campinas --- UNICAMP, Campinas, SP, Brazil}
\affil[3]{Universidade Federal do Maranhão --- UFMA, São Luís, MA, Brazil}

\date{May 2026}

\hypersetup{
  pdftitle={Margin Play -- MARL for Public Policy Analysis in the Brazilian Equatorial Margin},
  pdfauthor={A. S. Leitao Filho et al.},
  pdfkeywords={MARL, BRO, CTDE, Brazilian Equatorial Margin, Public Policy, Maranhao, royalties}
}

\begin{document}
\maketitle

\begin{abstract}
The Brazilian Equatorial Margin (BEM) constitutes the next frontier
of offshore oil exploration in Brazil, with operations expected
to begin in 2026 in the Foz do Amazonas basin.
The BEM assets are fiscally and territorially linked, primarily,
to the state of Maranhão, currently the state with the lowest
Human Development Index in the Federation (HDI 0.676, IBGE 2022).
This gives rise to the central public policy question of this
work: to what extent and under what conditions does oil exploration
in the BEM generate net positive externalities for the state of
Maranhão?
The problem is intrinsically multi-agent: the Federal Government
seeks tax revenue and energy security; the state seeks regional
welfare under constitutional royalty earmarking; the operator
maximizes profit subject to operational risk; the ANP and IBAMA
represent conflicting institutional mandates; and Amazonian
communities assign greater weight to territorial and environmental
vectors than to monetary income.
We present Margin Play, a Multi-Agent Reinforcement Learning (MARL)
system that simulates these tensions under Brazilian empirical
calibration (Law 9.478/97, Law 12.858/2013, Petrobras Form 20-F,
TCU 2.936/2021, MMA NDC 2024, CPT/INCRA/CIMI 2024) and grounded
in classical economic literature (Aschauer--Munnell, Cobb--Douglas,
CRRA, Atkinson, van der Ploeg).
The system implements six agents under the CTDE paradigm
(Centralized Training, Decentralized Execution) trained with
BRO-MARL.
The results, obtained from 60,000 episodes distributed across
six scenarios (three macroeconomic baselines, two policy
counterfactuals and one structural transformative regime designated
MA-Próspero), indicate that the answer to the central question is
conditional on the adopted institutional regime: under the reference
baseline, the welfare gain is marginal
($W_{\mathrm{aval}}\!\approx\!1.68$), whereas the MA-Próspero
parametric configuration yields $\Delta W = {+}17.5\%$ and
$\Delta R_{\mathrm{com}} = {+}21.3\%$, simultaneously with an
environmental liability below the reference
($E_{\mathrm{amb}} = 0.048$ vs.\ $0.076$).
It is concluded that the fundamental problem does not consist in
a trade-off between production and welfare, but in the choice of
the public policy regime linked to exploration.
\end{abstract}


\section{Introduction}
\label{sec:intro}

\subsection{Context and Motivation}
\label{subsec:context}

\begin{figure}[!ht]
  \centering
  \includegraphics[width=\linewidth]{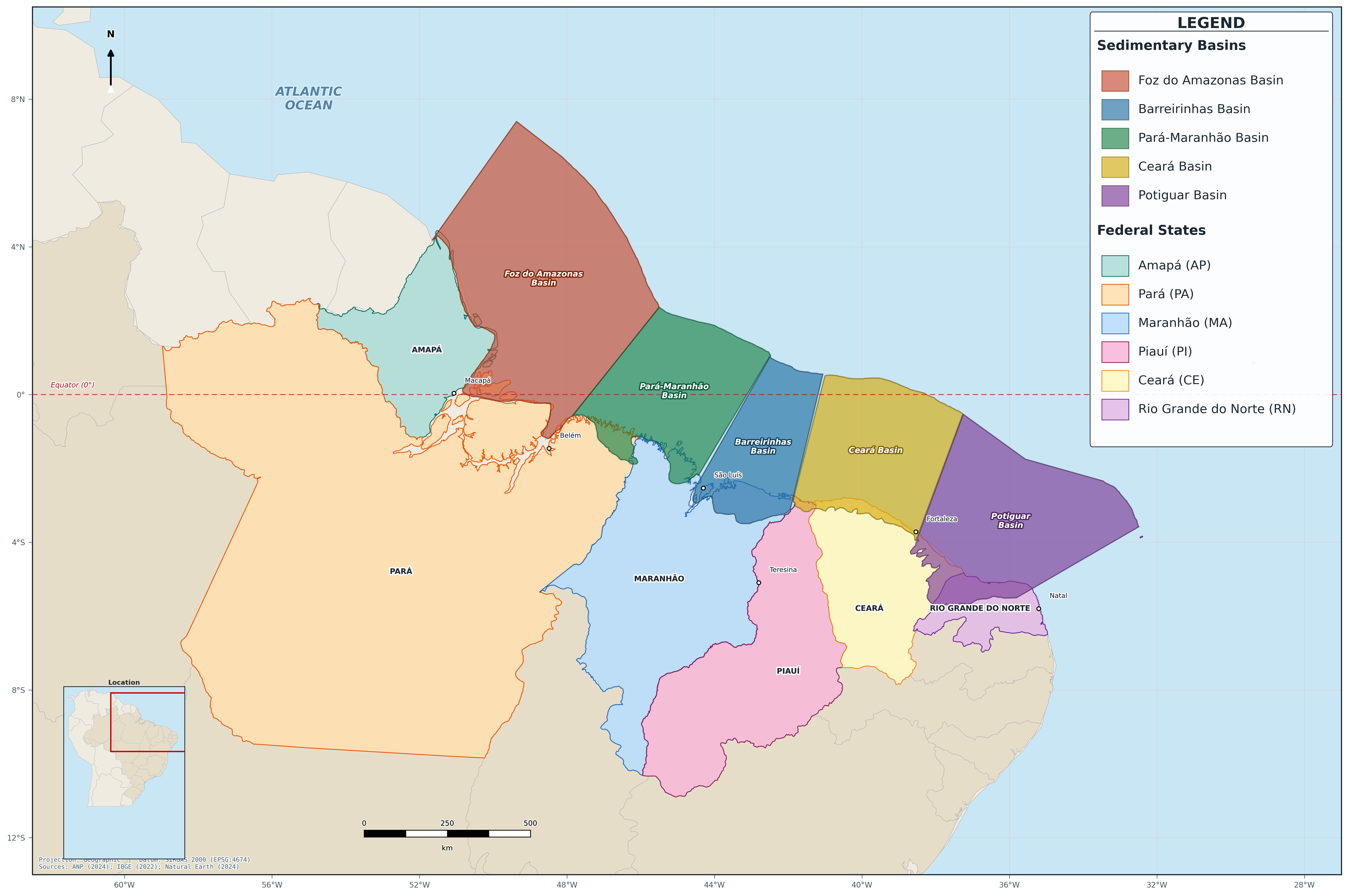}
  \caption{Map of the Brazilian Equatorial Margin (BEM) with the
    five offshore sedimentary basins (Foz do Amazonas,
    Pará-Maranhão, Barreirinhas, Ceará and Potiguar) and the
    state of Maranhão highlighted as the primary fiscal-territorial
    affiliate.
    \textit{The designations employed and the presentation of
    material in this map do not imply the expression of any
    opinion on the part of the authors concerning the legal status
    of any country, territory, city or area.}}
  \label{fig:bemmap}
\end{figure}

The Brazilian Equatorial Margin (BEM) is a geological complex
extending from Amapá to Rio Grande do Norte, encompassing five
offshore sedimentary basins (Foz do Amazonas, Pará-Maranhão,
Barreirinhas, Ceará and Potiguar),
as illustrated in \cref{fig:bemmap}.
Estimates of undiscovered potential from the National Petroleum,
Natural Gas and Biofuels Agency (ANP) suggest potential reserves
of 10 to 30 billion barrels of oil equivalent
\cite{anp2025meb}.
It is noteworthy that the Barreirinhas basin --- whose
fiscal-territorial affiliation also falls on the state of Maranhão
--- presents distinct techno-economic optimization challenges,
notably in carbonate matrix stimulation processes, whose viability
depends on the integration of reactive flow simulations with Net
Present Value (NPV) metrics, directly influencing the cost and
risk structures assumed by operators \cite{anp2025meb}.

The region became the focus of regulatory tension between 2023
and 2025 with the FZA-M-59 case: Petrobras applied for a drilling
licence in the Foz do Amazonas basin; the Brazilian Institute of
Environment and Renewable Natural Resources (IBAMA) denied the
licence \cite{ibama2023fza} due to deficiencies in the emergency
plan; the Ministry of Environment (MMA), initially aligned with
IBAMA, subsequently positioned itself in favour of re-analysis
\cite{mma2024ndc}.
The conflict illustrates the complexity of the multi-agent
problem: a single project involves fiscal interests (Federal
Government), constitutional equity interests (state), operational
interests (Petrobras), procedural-environmental interests (IBAMA),
technical-regulatory interests (ANP) and socio-territorial
interests (Amazonian communities), under legal constraints ranging
from the Federal Constitution of 1988 \cite{brasil1988cf} to
specific resolutions such as ANP Resolution 882/2022
\cite{anp2022res882}.

Following the granting of the drilling licence for the Morpho
FZA-M-59 well in October 2025, Petrobras announced an investment
of approximately US\$\,3 billion for the drilling of 16 wells in
the BEM between 2025 and 2029, consolidating the transition from
geological promise to operational reality.
This development validates the urgency of the analytical question
addressed in this study.

The central question of this work can be stated as: to what extent
and under what conditions does oil exploration in the BEM generate
net positive externalities for the state of Maranhão?
Maranhão is the state with primary fiscal-territorial affiliation
to the Foz do Amazonas and Pará-Maranhão basins, and the one with
the lowest HDI in the federation (0.676, IBGE 2022).
The resource curse literature \cite{vanderploeg2010,corden1982}
indicates that the influx of oil revenues produces ambiguous
effects: positive via royalties, FUNDEB and GDP; negative via
Dutch disease, institutional capture and intensification of
territorial conflicts.
The question is not purely theoretical: in 2024 the state
recorded 363 land conflicts (1st place nationally, CPT
\cite{cpt2024}) and 49 cases of violence against indigenous
peoples (CIMI \cite{cimi2024}).
Oil exploration in the BEM will take place within this
pre-existing socio-territorial context.

Moreover, Hodler, Lechner and Raschky (2023) \cite{hodler2023},
applying a causal forest estimator to data from 3,800 sub-Saharan
African districts, empirically confirmed that stronger institutions
amplify the positive effect of mining activities on economic
development and attenuate their negative effect on conflicts ---
providing causal machine learning validation for the institutional
conditionality that is the central premise of this work.

Public policy analysis for this scenario requires analytical tools
with four simultaneous properties: mandate heterogeneity,
physical-economic coupling, calibration with national data, and
robust counterfactual capacity
\cite{lucas1976,aschauer1989,petrobras2024,acemoglu2012,brasil2013lei12858}.

\subsection{Objectives and Contributions}
\label{subsec:contributions}

Margin Play is proposed as a computational instrument for public
policy analysis, and its contributions are organised along five
dimensions.
The first is a six-agent multi-agent architecture calibrated with
Brazilian primary sources, including updated territorial
calibration (H-TERR-2) based on CPT/INCRA/CIMI 2024 data
(\cref{sec:metodologia}).
The second consists of reward functions with heterogeneous
functional forms (CRRA, log-Aschauer, Atkinson, Cobb--Douglas,
Hubbert) and $R_{\mathrm{gov}}$ with decomposition of Net Current
Revenue into five components endogenously derived from Brazilian
fiscal legislation.
The third is the application of the BRO algorithm \cite{nauman2024bro}
under the CTDE paradigm \cite{lowe2017maddpg}, with distributional
critic TQC (Truncated Quantile Critics) \cite{kuznetsov2020tqc}.
The fourth is the counterfactual analysis across six scenarios
(\cref{sec:resultados}): three macroeconomic baselines, two policy
counterfactuals and one structural transformative regime
(MA-Próspero), totalling 60,000 episodes.
The fifth is the characterisation of the MA-Próspero regime --- a
parametric configuration of six simultaneous structural
interventions that, at equilibrium, yields $\Delta W = {+}17.5\%$
accompanied by an environmental liability below the reference.

\section{Methodological Rationale}
\label{sec:justificativa}

\subsection{MARL as a Public Policy Analysis Tool}
\label{subsec:marl-policy}

Traditional public policy analysis, anchored in Computable
General Equilibrium (CGE) models or reduced-form econometric
estimations, presents important limitations for the BEM case.
CGE models assume, by construction, representative agents with
homogeneous utility functions --- a premise incompatible with
the mandate heterogeneity among ANP, IBAMA and MMA
\cite{shoham2009}.
Reduced-form econometric models are essentially retrospective
and lose validity when applied to states not yet realised, such
as the BEM's full-production regime \cite{lucas1976}.

Multi-Agent Reinforcement Learning (MARL) simultaneously overcomes
these limitations: it admits agents with arbitrary objective
functions, allows the learning of approximate Nash equilibria
through repeated interaction \cite{lowe2017maddpg}, and the
evaluation of trade-offs is performed directly via Monte Carlo
sampling.

Margin Play's positioning in the literature requires explicit
differentiation.
The AI Economist \cite{zheng2022} established a two-level MARL
framework for fiscal policy design.
Margin Play extends this paradigm in three directions:
(i)~from two hierarchical levels to six institutionally
heterogeneous agents, each with distinct and legally grounded
objective functions;
(ii)~from a generic economy to a specific subnational
fiscal-territorial context, calibrated with Brazilian primary data;
and (iii)~from symmetric mandates to structurally conflicting
mandates --- the IBAMA reward function is fundamentally opposed
to that of the operator.

In the petroleum domain, \citet{radovic2022} employed deep MARL
(A2C + PPO) to reveal robust macro strategies of international
oil companies across 408 energy transition scenarios.
Margin Play is complementary: where \citet{radovic2022} examine
strategies of competing corporations, Margin Play examines
equilibrium dynamics from the perspective of the institutional
system surrounding the operator.
In the domain of computationally calibrated simulation for Brazil,
PolicySpace2 \cite{furtado2022} constitutes the most relevant
national precedent, which Margin Play advances by replacing fixed
behavioural rules with strategic learning.

\subsection{Centralized Training and Decentralized Execution
  (CTDE)}
\label{subsec:ctde}

The CTDE paradigm \cite{lowe2017maddpg} is particularly
appropriate for the problem considered here.
During centralised training, each critic observes the global
system state, which mitigates the non-stationarity typical of
purely decentralised MARL regimes.
During execution, each actor decides based exclusively on its
own local observation, reproducing the Brazilian institutional
reality.

Margin Play implements CTDE with BRO-MARL, a combination that
provides:
(a)~centralised critic access to the global state during
training;
(b)~distributional TQC critics with 25 quantile atoms,
addressing high-variance return regimes; and
(c)~six structurally heterogeneous actor networks, each with
reward functions specific to its institutional mandate.

\subsection{Brazilian Empirical Calibration}
\label{subsec:calibration}

Margin Play is designed as an analytical tool for the specific
Brazilian case, not for a generic oil-producing country.
The mandatory earmarking of royalties to health and education,
in the proportion of 75\%/25\%, is treated as a hard constraint
of the physical model, as stipulated by Law 12.858/2013
\cite{brasil2013lei12858}, and not as an agent decision.
The Federal Government's share of offshore royalties is
parametrised at approximately 22\%, estimated from the fiscal
regime in force under Law 9.478/1997 \cite{brasil1997lei9478},
maintained in force since March 2013 by an injunction granted
in ADI 4917 MC/DF \cite{stf2018adi4917}.
The operator's lifting cost is calibrated from Petrobras Form
20-F 2024 \cite{petrobras2024}, not from OPEC averages.
The zonal weights of the Community agent's reward function
reflect Brazilian Amazonian ethnography, as per Escobar
\cite{escobar2008}, Almeida \cite{almeida2008}, CIMI
\cite{cimi2024} and CPT \cite{cpt2024}.

\section{Methodology}
\label{sec:metodologia}

\subsection{System Overview}
\label{subsec:overview}

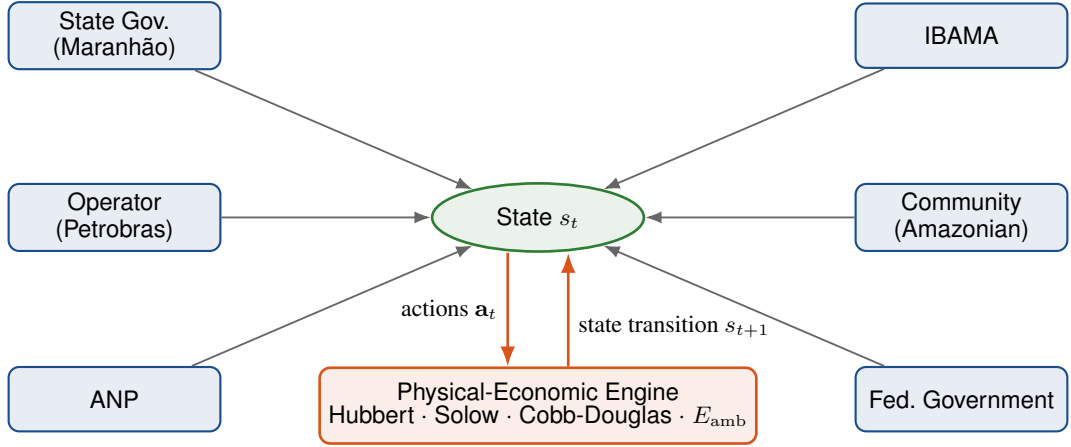
\begin{figure*}[htb]
  \centering
  \resizebox{0.85\linewidth}{!}{%
  \begin{tikzpicture}[
      node distance=0.75cm and 0.9cm,
      every node/.style={font=\small\sffamily},
      agent/.style={
          rectangle, rounded corners=4pt, draw=rulecol, line width=0.8pt,
          fill=rulecol!10, minimum width=2.8cm, minimum height=0.85cm,
          align=center},
      physmod/.style={
          rectangle, rounded corners=4pt, draw=accent1, line width=1pt,
          fill=accent1!10, minimum width=4.2cm, minimum height=1.0cm,
          align=center},
      state/.style={
          ellipse, draw=accent2, line width=1pt, fill=accent2!10,
          minimum width=2.8cm, minimum height=0.9cm, align=center},
      arrow/.style={-Latex, line width=0.7pt, draw=neutral}
  ]
  \node[state] (state) at (0,0) {State $s_t$};
  \node[agent] (gov)   at (-5.6,  2.4) {State Gov.\\(Maranhão)};
  \node[agent] (oper)  at (-5.6,  0.0) {Operator\\(Petrobras)};
  \node[agent] (anp)   at (-5.6, -2.4) {ANP};
  \node[agent] (ibama) at ( 5.6,  2.4) {IBAMA};
  \node[agent] (com)   at ( 5.6,  0.0) {Community\\(Amazonian)};
  \node[agent] (fed)   at ( 5.6, -2.4) {Fed.\ Government};

  \foreach \src in {gov, oper, anp, ibama, com, fed}
      \draw[arrow] (\src) -- (state);

  \node[physmod, below=1.5cm of state] (phys) {
      Physical-Economic Engine\\[-1pt]
      {\footnotesize Hubbert $\cdot$ Solow $\cdot$ Cobb-Douglas $\cdot$ $E_\mathrm{amb}$}
  };

  \draw[arrow, line width=1pt, draw=accent1] ([xshift=-0.4cm]state.south) -- node[midway, left, font=\footnotesize] {actions $\mathbf{a}_t$} ([xshift=-0.4cm]phys.north);
  \draw[arrow, line width=1pt, draw=accent1] ([xshift=0.4cm]phys.north) -- node[pos=0.35, right, font=\footnotesize] {state transition $s_{t+1}$} ([xshift=0.4cm]state.south);

  \end{tikzpicture}}%
  \caption{System architecture of the Margin Play framework under the CTDE paradigm. Six institutional agents interact with the physical-economic environment. During centralized training, critics observe the full global state $s_t$; during execution, each actor conditions its policy strictly on its local observation space.}
  \label{fig:architecture}
\end{figure*}

\Cref{fig:architecture} presents the architecture of Margin Play.
The fundamental cycle chains five steps: each of the six agents
observes the global state $s_t$ and produces an action $a^i_t$;
the combined actions $\mathbf{a}_t = (a^1_t, \ldots, a^6_t)$
feed the physical-economic module; the state evolves from $s_t$
to $s_{t+1}$ via transition equations (Hubbert, Solow,
Cobb--Douglas, environmental liability dynamics); each agent
receives a specific reward $r^i_t$ and stores the tuple in the
replay buffer; and BRO-MARL training updates actors and critics
under the CTDE paradigm.

\subsection{World State and Operational Sequence}
\label{subsec:state}

The state $s_t$ contains physical, economic and institutional
variables (\cref{tab:estado}).
The zonal variables $K_{\mathrm{pub}}$, $K_{\mathrm{hum}}$,
$K_{\mathrm{sa\acute{u}de}}$, $K_{\mathrm{priv}}$,
$C_{\mathrm{inst}}$ and $N$ are 4-dimensional vectors, with one
entry per mesoregion of Maranhão according to the IBGE
classification.

\begin{table}[!ht]
\centering
\caption{World state variables $s_t$.}
\label{tab:estado}
\small
\begin{tabular}{@{}lll@{}}
\toprule
\textbf{Variable} & \textbf{Type} & \textbf{Description} \\
\midrule
$P_{\mathrm{eff}}$  & scalar & Effective production (Mbbls/d) \\
$R$                 & scalar & Remaining reserves (Gbbl) \\
$\mathit{price}$    & scalar & Brent price (USD/bbl) \\
$E_{\mathrm{amb}}$  & scalar $\in [0,1]$ & Environmental liability \\
$K_{\mathrm{pub}}$  & 4D vector & Public capital by zone \\
$K_{\mathrm{hum}}$  & 4D vector & Human capital (education) by zone \\
$K_{\mathrm{sa\acute{u}de}}$ & 4D vector & Health capital by zone \\
$K_{\mathrm{priv}}$ & 4D vector & Private capital by zone \\
$C_{\mathrm{inst}}$ & 4D vector $\in [0,1]$ & Institutional capital \\
$N$                 & 4D vector & Population by zone (thousands) \\
ICMS, FPE, FUNDEB   & scalars & State revenues (R\$\,bn) \\
$\mathit{GDP}_{\mathrm{state}}$ & scalar & State GDP (R\$\,bn) \\
$W$                 & scalar & Regional aggregate welfare \\
\bottomrule
\end{tabular}
\end{table}

\begin{figure}[!ht]
  \centering
  \includegraphics[width=0.6\linewidth]{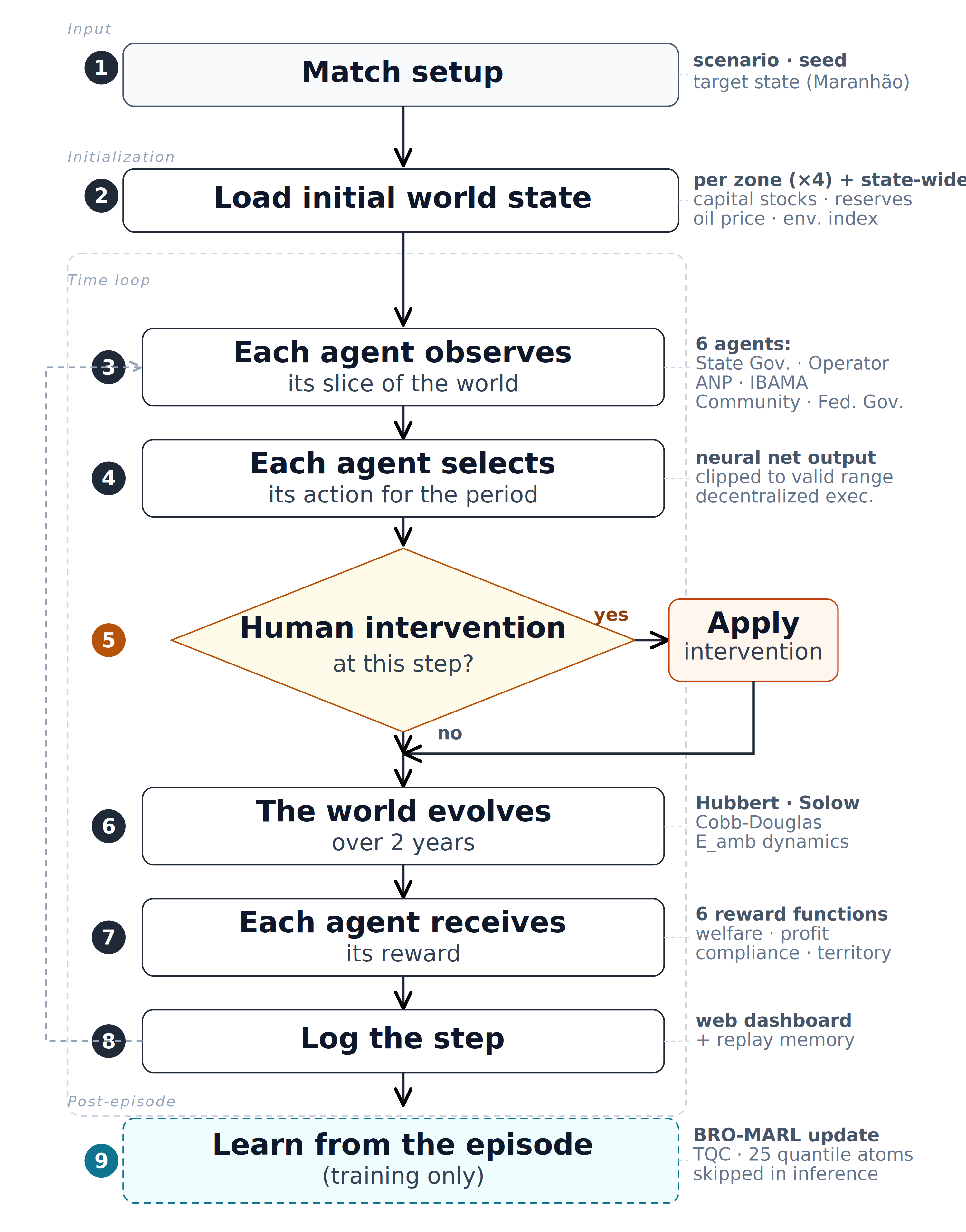}
  \caption{Operational sequence of a Margin Play episode.
    Input stage (steps 1--2), 15-step biennial time loop
    (steps 3--8) and post-episode with BRO-MARL update
    (step 9, suppressed in inference mode).}
  \label{fig:flowchart}
\end{figure}

\Cref{fig:flowchart} details the operational sequence of a
complete episode, organised in three stages.
In the input stage (step~1), the episode is initialised by the
specification of a scenario, a random seed and the target
state of Maranhão.
In the initialisation stage (step~2), the state $s_0$ is loaded
with the initial conditions of the four mesoregions --- capital
stocks, reserves, Brent price and initial environmental liability.

The time loop (steps~3--8) constitutes the core of the episode
and repeats 15 biennial iterations, totalling a 30-year horizon.
In step~3, each agent observes exclusively the fraction of the
global state accessible under its mandate, in accordance with
the decentralised execution of CTDE: the State Government
observes capital stocks, sovereign fund, Gini index and
ICMS/FPE revenues; the Operator observes reserves, price,
environmental liability and inspection intensity; the ANP
observes production, royalties, operational safety and local
content; IBAMA observes environmental liability, community
mobilisation and accountability metrics; the Community observes
income, environmental quality, employment and human capital;
the Federal Government observes national production, total
royalties, environmental liability and price.
In step~4, each agent produces its action from the neural
network output, clipped to the valid range of each dimension.
Step~5 reserves an optional human intervention point; in
standard training and inference mode, it is ignored.
In step~6, the world evolves two years through the
physical-economic transition equations.
In step~7, each agent receives its individual reward $r^i_t$
and the tuple is stored in the replay buffer.
Step~8 records the state in the visualisation panel (observer
mode) or in the replay memory (training mode); upon reaching
step~14, the episode ends.
In the post-episode stage (step~9), in training mode,
BRO-MARL samples 256 transitions and updates the six actors
and the six distributional TQC critics; in inference mode,
this step is suppressed.

\FloatBarrier

\subsection{Physical-Economic Model}
\label{subsec:phys-econ}

\subsubsection{Oil Production}

Effective oil production is modelled by the asymmetric Hubbert
curve \cite{hubbert1956,brandt2010}, a classical parametrisation
to represent the temporal trajectory of fields with ramp-up,
peak and progressive decline phases:
\begin{equation}
P_t^{\mathrm{ef}} = P_{\max}^{\mathrm{ef}} \cdot
  e^{a (t/t_{\mathrm{peak}} - 1)} \cdot
  (t/t_{\mathrm{peak}}) \cdot
  \alpha_{\mathrm{inv}} \cdot
  \mathbf{1}_{R_t > 0}.
\label{eq:hubbert}
\end{equation}
Here $P_{\max}^{\mathrm{ef}}$ denotes the maximum effective
production (in Mbbls/d), $t_{\mathrm{peak}}$ the peak instant
and $a$ the asymmetry coefficient.
The adoption of the asymmetric form, rather than the classical
symmetric Hubbert, reflects the empirical observation that
modern fields tend to exhibit faster decline than ascent, due
to secondary and tertiary recovery technologies.
The multiplication by $\alpha_{\mathrm{inv}} \in [0,1]$ couples
the curve to the operator's investment action; the indicator
$\mathbf{1}_{R_t>0}$ ensures that production ceases exactly
when reserves are exhausted.

The maximum effective capacity results from the multiplicative
composition of two regulatory levers:
$P_{\max}^{\mathrm{ef}} = P_{\max} \cdot
\kappa_{\mathrm{auction}} \cdot \kappa_{\mathrm{aprov}}$,
where $\kappa_{\mathrm{auction}} \in [1;\, 1.2]$ is determined
by the offshore block licensing pace controlled by the Federal
Government and captures the speed at which new blocks are
auctioned in ANP rounds.
The factor $\kappa_{\mathrm{aprov}} \in [0.8;\, 1.2]$ is
determined by the ANP's speed of approval of Development Plans
and explicitly models the regulatory bottleneck of Development
Plan approval, whose observed average lag in the BEM is of the
order of 18 months, according to a TCU operational audit
\cite{tcu2021}.

\subsubsection{Environmental Liability Dynamics}

The temporal trajectory of environmental liability
$E_{\mathrm{amb}}$ is described by a balance equation with four
terms, each representing a distinct physical-economic mechanism:
\begin{equation}
\frac{\Delta E_{\mathrm{amb}}}{\Delta t} =
  \alpha_{\mathrm{base}} P +
  \alpha_{\mathrm{prod}}^{\mathrm{ef}} P (1 - \varphi) -
  \theta_{\mathrm{rem}} I_{\mathrm{amb}} E_{\mathrm{amb}} -
  \gamma_{\mathrm{nat}} E_{\mathrm{amb}}.
\label{eq:eamb}
\end{equation}
The term $\alpha_{\mathrm{base}} P$ represents the irreducible
environmental impact associated with production --- liabilities
that persist even under perfect inspection: water-cut, discarded
drilling fluids, fugitive emissions.%
\footnote{Fugitive emissions are quantified in accordance with
  the 2019 Refinement to the 2006 IPCC Guidelines, Volume~2,
  Chapter~4; supplementary reference: CONAMA Resolution 462/2014.}
The term $\alpha_{\mathrm{prod}}^{\mathrm{ef}} P (1 - \varphi)$
represents the avoidable impact, linearly attenuated by the local
environmental inspection intensity $\varphi \in [0,1]$;
additionally,
$\alpha_{\mathrm{prod}}^{\mathrm{ef}} = \alpha_{\mathrm{prod}}
\cdot (1 - 0.4 \cdot \text{fed\_env\_insp})$
incorporates the influence of the federal environmental inspection
budget (IBAMA/ICMBio, per the 2024--2027 PPA), establishing a
direct channel between the Federal Government's environmental
inspection action and environmental dynamics.
The term $-\theta_{\mathrm{rem}} I_{\mathrm{amb}} E_{\mathrm{amb}}$
represents active remediation proportional to the existing
liability, encoding the principle that there is no remediation
where there is no liability.
Finally, $-\gamma_{\mathrm{nat}} E_{\mathrm{amb}}$ represents
natural decay through abiotic processes.

\subsubsection{Capital Stocks}

The five capital stocks relevant to regional welfare follow the
Solow accumulation structure, combined with the Aschauer--Munnell
foundation for public capital:
\begin{equation}
K_{i,t+1} = (1 - \delta_i)\, K_{i,t} + \eta_i \cdot I_{i,t},
\label{eq:capital}
\end{equation}
where $\delta_i$ is the annual depreciation rate of stock $i$
and $\eta_i$ is the investment-to-installed-capital conversion
efficiency.
\Cref{tab:capitais} reports the calibrated values, all anchored
in established academic sources.
The heterogeneity among stocks is methodologically significant:
$K_{\mathrm{pub}}$ depreciates at 4\% per year with an elasticity
of 0.39 in GDP, reflecting the centrality of public capital in
middle-income economies \cite{aschauer1989,munnell1990};
$K_{\mathrm{hum}}$ exhibits lower depreciation (2\%) consistent
with Lucas's hypothesis \cite{lucas1988} of increasing returns
to knowledge; $K_{\mathrm{sa\acute{u}de}}$ depreciates at 5\%
following Grossman \cite{grossman1972}; and $K_{\mathrm{priv}}$
at 7\%, standard Cobb--Douglas calibration \cite{cobb1928}.

\begin{table}[!ht]
\centering
\caption{Capital stock calibration.}
\label{tab:capitais}
\small
\begin{tabular}{@{}llll@{}}
\toprule
\textbf{Stock} & $\delta$ \textbf{(p.a.)} & \textbf{GDP Elast.} & \textbf{Source} \\
\midrule
$K_{\mathrm{pub}}$   & 4\% & $\beta \approx 0.39$
  & \citeauthor{aschauer1989}~\cite{aschauer1989},
    \citeauthor{munnell1990}~\cite{munnell1990} \\
$K_{\mathrm{hum}}$   & 2\% & $\gamma \approx 0.25$
  & \citeauthor{glomm1992}~\cite{glomm1992},
    \citeauthor{lucas1988}~\cite{lucas1988} \\
$K_{\mathrm{sa\acute{u}de}}$ & 5\% & $\delta \approx 0.15$
  & \citeauthor{grossman1972}~\cite{grossman1972} \\
$K_{\mathrm{priv}}$  & 7\% & $\alpha \approx 0.30$
  & \citeauthor{cobb1928}~\cite{cobb1928} \\
\bottomrule
\end{tabular}
\end{table}

The dynamics of institutional capital deviate from the Solow
standard by incorporating a rent capture term, following the
institutional economics literature
\cite{acemoglu2012,north1990,mendes2014}:
\begin{equation}
C_{\mathrm{inst},t+1} =
  C_{\mathrm{inst},t} +
  \gamma_{\mathrm{learn}} \cdot I_{\mathrm{inst},t} -
  \gamma_{\mathrm{cap}} \cdot \rho_t \cdot
  \mathbf{1}_{C_{\mathrm{inst},t}>0}.
\label{eq:cinst}
\end{equation}
The first motion term captures the institutional learning
resulting from investment in technical and regulatory capacity.
The second, $-\gamma_{\mathrm{cap}} \cdot \rho_t$, explicitly
models the deterioration associated with oil rent capture.
The parameter $\gamma_{\mathrm{cap}} = 0.02$ is calibrated so
that $\rho \approx 0.5$ reduces $C_{\mathrm{inst}}$ by
approximately 10 percentage points per biennial step, a magnitude
consistent with longitudinal observations of Brazilian
municipalities subject to oil boom cycles, notably Mossoró/RN
and Macaé/RJ.

\subsubsection{State Net Current Revenue}

Net Current Revenue is endogenously decomposed into five
components, each derived from Brazilian fiscal legislation:
\begin{equation}
\mathrm{NCR} =
  \mathrm{ICMS}_{\mathrm{end}} +
  \mathrm{FPE} +
  \mathrm{FUNDEB} +
  \mathrm{Other} +
  \mathrm{Royalties}_{\mathrm{earmarked}}.
\label{eq:rcl}
\end{equation}
The component $\mathrm{ICMS}_{\mathrm{end}}$ is endogenised to
state GDP dynamics via
$\mathrm{ICMS}_{\mathrm{end}} = \varepsilon \cdot
\mathit{GDP}^{\varepsilon}_{\mathrm{state}}$,
with $\varepsilon$ calibrated from the ICMS-GDP elasticity series
of STN-IPEA \cite{stn2024finbras,ipea2024}.
The State Participation Fund (FPE) is proportional to the national
pool of income tax and IPI according to the coefficients of
Complementary Law 91/97 \cite{brasil1997lc91}.
FUNDEB follows Constitutional Amendment 108/2020
\cite{brasil2020ec108} and is entirely earmarked for education.
State royalties enter the model already earmarked by Law
12.858/2013 \cite{brasil2013lei12858}, with the 75\%/25\%
allocation to health and education treated as a hard constraint
of the physical model --- not as a decision of the State
Government agent --- which reflects the statutory nature of the
earmarking in the Brazilian legal order.

\subsubsection{State GDP with Five Factors}

Aggregate state production is modelled by a five-factor
Cobb--Douglas function:
\begin{equation}
\mathit{GDP}_{\mathrm{state},t+1} =
  A \cdot K_{\mathrm{priv},t}^{\alpha} \cdot
  K_{\mathrm{pub},t}^{\beta} \cdot
  K_{\mathrm{hum},t}^{\gamma} \cdot
  K_{\mathrm{sa\acute{u}de},t}^{\delta} \cdot
  N_t^{\eta}.
\label{eq:pib}
\end{equation}
The simultaneous inclusion of $K_{\mathrm{sa\acute{u}de}}$ and
$K_{\mathrm{hum}}$ as distinct productive factors is
methodologically essential: it allows the separate estimation of
the macroeconomic effects of the two royalty earmarkings
established by Law 12.858/2013 \cite{brasil2013lei12858},
enabling counterfactual analysis of the revocation or extension
of this earmarking.

\FloatBarrier

\subsection{The Six Agents}
\label{subsec:agents}

Each Margin Play agent has a distinct institutional mandate, its
own action space and a reward function calibrated to its legal
obligations.
The heterogeneity of these functions is the central element of
the architecture: the IBAMA agent and the Operator agent respond
to structurally opposing incentives, reproducing the real tensions
of the FZA-M-59 case.

\subsubsection{State Government (Maranhão)}

The State Government's mandate is to maximise regional welfare
under the constraints of the Fiscal Responsibility Law and
constitutional spending earmarks.
In each biennial period, the agent decides the allocation of six
normalised budgetary fractions among free education, free health,
infrastructure, institutional strengthening, sovereign fund
constitution and interior distribution.
The reward function $R_{\mathrm{gov}}$ combines eight weighted
components:
\begin{equation}
\begin{split}
R_{\mathrm{gov}} = 0.10\,\Delta W + 0.12\,u_{\mathrm{edu}} + 0.12\,u_{\mathrm{sa\acute{u}de}} + 0.20\,u_{\mathrm{infra}} \\
+ 0.10\,u_{\mathrm{inst}} + 0.07\,b_{\mathrm{fund}} + 0.10\,u_{\mathrm{equid}} + 0.20\,\mathit{visib} - 0.10\,\mathit{def}_{\mathrm{pen}}
\end{split}
\label{eq:rgov}
\end{equation}
The functional forms are heterogeneous, reflecting the distinct
nature of each objective: CRRA utility \cite{pratt1964} for
education and health, log-Aschauer for infrastructure, zonal
Atkinson index \cite{atkinson1970} for distributive equity, and
log-visible variation for the electoral cycle component
\cite{drazen2010,rogoff1990,persson2000}.

\subsubsection{Operator (Petrobras)}

The Operator pursues the mandate of maximising accounting profit,
controlling two decision variables: the investment intensity in
production $\alpha_{\mathrm{inv}}$ and the effort dedicated to
operational safety $\alpha_{\mathrm{seg}}$.
The reward function $R_{\mathrm{oper}}$ is:
\begin{equation}
R_{\mathrm{oper}} = \mathit{rev} - \mathit{cost}_{\mathrm{op}} - \mathit{cost}_{\mathrm{sec}} - \mathit{cost}_{\mathrm{reg}} - p_{\mathrm{acc}} L \cdot \mathit{rev} - \mathit{risk}_{\mathrm{reg}}
\label{eq:roper}
\end{equation}
Operational cost is calibrated from the Petrobras 2024 Form 20-F
($\mathit{cost}_{\mathrm{op}} = 0.08 \cdot P_{\mathrm{eff}} \cdot
\mathit{price}$ \cite{petrobras2024}), security cost adopts a
quadratic structure \cite{bier2004}, and the component
$\mathit{risk}_{\mathrm{reg}}$ is activated when
$E_{\mathrm{amb}} > 0.20$, reproducing the liability precedent
of the Frade-Chevron case \cite{pgr2011frade,deepwater2011}.

\subsubsection{ANP --- National Petroleum Agency}

The ANP operates with a dual mandate: to maximise institutional
revenue from production and to ensure operational safety of
fields, in compliance with Law 9.478/97 \cite{brasil1997lei9478}
and TCU determinations \cite{tcu2021}.
The ANP controls the speed of Development Plan approvals ---
whose observed average delay in the BEM is of the order of 18
months \cite{tcu2021} --- and the rigour applied in operational
safety audits.
The reward function $R_{\mathrm{anp}}$ is:
\begin{equation}
R_{\mathrm{anp}} = {+}0.10\,\mathit{rev}_{\mathrm{state}} + 0.45\,\mathit{op\_sec} - 0.20\,\Delta E_{\mathrm{amb}} - 0.15\,\mathit{eco\_dmg} - 0.15\,\varphi_{\mathrm{insp}} \cdot \mathit{pressure}
\label{eq:ranp}
\end{equation}

\subsubsection{IBAMA}

IBAMA operates with the mandate of ensuring the procedural-environmental
compliance of projects, in accordance with the National
Environmental Policy (Law 6.938/81 \cite{brasil1981lei6938}).
The IBAMA agent controls the environmental inspection intensity
$\varphi_{\mathrm{insp}}$ and the degree of environmental
compensation required of the operator.
The reward function $R_{\mathrm{ibama}}$:
\begin{equation}
R_{\mathrm{ibama}} = {-}0.55\,\Delta E_{\mathrm{amb}} - 0.30\,\mathit{eco\_dmg} + 0.15\,\mathit{accountability} - 0.10\,\varphi_{\mathrm{insp}} \cdot \mathit{pressure}
\label{eq:ribama}
\end{equation}
strongly penalises environmental liability accumulation, with a
dominant weight of $0.55$, in line with the mandate empirically
observed in the FZA-M-59 case \cite{ibama2023fza}.

\subsubsection{Community (H-TERR-2 Territorial Calibration)}

The Community agent represents the populations of the four
mesoregions of Maranhão, with the mandate of maximising aggregate
local welfare by zone, weighted by coefficients calibrated to
Brazilian Amazonian ethnography and primary 2024 territorial data.
In each period, the Community agent chooses the level of
territorial mobilisation and declares its relative environmental
preference.
The reward function $R_{\mathrm{com}}$ weights zones by
population share:
\begin{equation}
R_{\mathrm{com}} =
  \sum_{z=1}^{4} \frac{N_z}{N_{\mathrm{total}}}\, r_z -
  0.15\,\mathrm{Gini} -
  \mathit{mob\_cost}
\label{eq:rcom}
\end{equation}
where the zonal welfare $r_z$ disaggregates the local determinants:
\begin{equation}
\begin{split}
r_z = 0.20\,\Delta\!\log(\mathit{income}_z) + 0.15\,\Delta K_{\mathrm{pub},z} + 0.20\,\Delta K_{\mathrm{sa\acute{u}de},z} \\
+ 0.25\,(1 - E_{\mathrm{amb}}) + 0.10\,\mathit{terr}_z - 0.10\,\max\!\bigl(0,\, \Delta n_z - 5\%\bigr)
\end{split}
\label{eq:rz}
\end{equation}
The component $\mathit{terr}_z$ (H-TERR-2 calibration) replaces
the previous environmental proxy with weighted primary data:
363 land conflicts in MA-2024 (CPT \cite{cpt2024}, 1st nationally),
49 cases of indigenous violence (CIMI \cite{cimi2024}), and INCRA
land titling bottleneck (3 titles / 424 pending processes
\cite{incra2024sipra,terradedireitos2024}).

\subsubsection{Federal Government}

The Federal Government pursues three simultaneous objectives:
to maximise federal revenue from production, to ensure national
energy security and to sustain Brazil's climate soft power in
international negotiations.
The Federal Government controls the offshore block licensing
pace, the rate of the Energy Domain Economic Intervention
Contribution on fuels ($\alpha_{\mathrm{CIDE}}$) and the federal
budget for environmental inspection (IBAMA/ICMBio).
The reward function $R_{\mathrm{fed}}$ is:
\begin{equation}
R_{\mathrm{fed}} = {+}0.35\,\mathit{rev}_{\mathrm{Union}} + 0.15\,\mathit{energy\_sec} - 0.30\,E_{\mathrm{amb}} + 0.20\,W - 0.05\,\mathit{env\_insp\_cost}
\label{eq:rfed}
\end{equation}

\subsection{Learning Algorithm: BRO-MARL}
\label{subsec:bro}

Agent training is performed through an adaptation of the
Bigger, Regularized, Optimistic (BRO) algorithm
\cite{nauman2024bro} to the multi-agent CTDE paradigm,
designated BRO-MARL.
Rather than directly approximating the action-value function
$Q(s,a)$ by a deterministic network, each critic estimates the
quantile distribution $Z(s,a)$ via TQC \cite{kuznetsov2020tqc},
with 25 atoms trained under Huber loss ($\kappa = 1.0$).
This choice is critical for numerical stability: high return
variance regimes --- such as the optimistic scenario, in which
state revenue scales by an order of magnitude --- tend to induce
Q-value explosion in deterministic critics, a phenomenon
naturally controlled by the distributional critic
\cite{bellemare2017distributional}.

Target network updates follow the Polyak scheme
\cite{lillicrap2016ddpg,fujimoto2018td3} with $\tau = 0.005$.
The replay buffer has capacity $10^5$ transitions and provides
batches of 256 samples per iteration.
Actor and critic learning rates are fixed at $3 \times 10^{-4}$.
The temporal discount factor is $\gamma = 0.95$, a choice that
prioritises convergence stabilisation over a finite horizon
without compromising the representation of intertemporal
trade-offs relevant to the problem.
The first five episodes of each scenario are executed in warm-up
mode, without network updates, in order to populate the replay
buffer with representative transitions before learning begins.

\subsection{Experimental Scenarios}
\label{subsec:scenarios}

The experimental set comprises six qualitatively distinct
configurations, summarised in \cref{tab:cenarios}.
The first three constitute macroeconomic baselines --- pessimistic,
reference and optimistic --- in which maximum productive capacity
($P_{\max}$), annual discovery probability ($\pi_{\mathrm{disc}}$)
and the Brent price vary simultaneously, establishing a plausible
range of behaviours under different oil fundamentals regimes.

\begin{table}[!ht]
\centering
\caption{Macroeconomic and structural parameters by scenario.}
\label{tab:cenarios}
\small
\begin{tabular}{@{}lcccc@{}}
\toprule
\textbf{Scenario} & $P_{\max}$ & $\pi_{\mathrm{disc}}$
  & \textbf{Brent} & \textbf{Characteristic} \\
 & \textbf{(Mbbls/d)} & & \textbf{(US\$)} & \\
\midrule
Pessimistic          & 5.0  & 0.05 & 50 & Low baseline \\
Reference            & 8.0  & 0.10 & 70 & Central baseline \\
Optimistic           & 12.0 & 0.15 & 90 & High baseline \\
Without Law 12.858   & 8.0  & 0.10 & 70 & Unearmarked royalties \\
Brent Shock          & 8.0  & 0.10 & 70 & $\sigma +$ negative drift \\
\textbf{MA-Próspero} & 8.0  & 0.10 & 70
  & 6 structural levers \\
\bottomrule
\end{tabular}
\end{table}

The two following scenarios are policy counterfactuals.
The \emph{Without Law 12.858/2013} scenario suppresses the
mandatory royalty earmarking to health and education and
redistributes the resulting revenue to other revenues, freely
allocated by the State Government agent.
The \emph{Brent Shock} scenario preserves the reference baseline
fundamentals, but amplifies price volatility
($\sigma_{\mathrm{price}}\colon 0.10 \to 0.25$) and introduces
a negative drift ($\mu_{\mathrm{drift}}\colon 0 \to -0.01$),
simulating a declining and volatile price regime typical of
accelerated energy transition.

The sixth scenario, \emph{MA-Próspero}, is the central object of
this study --- a structural transformative regime, defined by the
simultaneous application of six parametric interventions over the
reference baseline (\cref{tab:alavancas}).
The joint reading of \cref{tab:cenarios} allows the isolation of
what effectively varies in each experiment: the three baselines
exclusively alter the triad $(P_{\max}, \pi_{\mathrm{disc}},
\text{Brent})$; the two policy counterfactuals preserve this
triad at the reference level and modify, respectively, the
earmarking rule and the stochastic price regime; and the
MA-Próspero regime shares with the counterfactuals the fixing
of macroeconomic fundamentals at the central baseline, differing
through the simultaneous application of six structural
interventions.
This specification enables the methodological separation between
fundamentals effects and regime effects.

The choice of applying the six levers simultaneously, rather
than testing them in isolation, follows a methodological choice:
the resource curse literature
\cite{vanderploeg2010,acemoglu2012,vanderploeg2011} indicates
that point reforms frequently fail because the mechanisms of
institutional capture, Dutch disease and territorial conflict
are interdependent.
MA-Próspero is therefore formulated as a coherent institutional
hypothesis, and not as \textit{a posteriori} parametric
adjustment.

\begin{table}[!ht]
\centering
\caption{The six levers of the MA-Próspero scenario.}
\label{tab:alavancas}
\small
\begin{tabularx}{\linewidth}{@{}lXX@{}}
\toprule
\textbf{Lever} & \textbf{Parameter} & \textbf{Rationale} \\
\midrule
1.\ Maximum fiscal capture
  & $\tau_{\mathrm{royalty}}\colon 0.18 \to 0.25$
  & Ceiling Law 9.478/97 art.\ 47 \cite{brasil1997lei9478} \\
2.\ Mature operations
  & $\lambda_{\mathrm{acid}}\colon 0.030 \to 0.015$
  & Statoil/Equinor post-1995 (benchmark) \\
3.\ Enhanced earmarking
  & Other rev.\ earmarked: $0\% \to 50\%$
  & Hypothetical statutory enhancement \\
4.\ Initial institutional investment
  & $C_{\mathrm{inst}}^{(0)} \times 5\;
    ({\approx}0.05 \to {\approx}0.30)$
  & Learning-state agent from $t=0$ \cite{north1990} \\
5.\ Territorial regularisation
  & $\delta_{\mathrm{pressure}}\colon 0.03 \to 0.06$;
    $\lambda_{\mathrm{accum}}\colon 0.4 \to 0.2$
  & CPT-INCRA-CIMI
    \cite{cpt2024,cimi2024,terradedireitos2024} \\
6.\ Macroeconomic stability
  & $\sigma_{\mathrm{price}}\colon 0.20 \to 0.15$;
    $\mu_{\mathrm{drift}}\colon {-}0.005 \to {+}0.005$
  & Stable Brent + moderate growth \\
\bottomrule
\end{tabularx}
\end{table}

\section{Experimental Results}
\label{sec:resultados}

The results are obtained from 60,000 episodes (10,000 in each
of the six scenarios), executed on Apple Silicon hardware under
the MLX framework.
Total execution time was 378.3 min ($\approx 63.1$ min per
scenario).
Results are analysed along three dimensions:
(i)~phenomenological coherence of the system,
(ii)~policy counterfactual responsiveness, and
(iii)~quantitative viability of the MA-Próspero regime as an
alternative institutional configuration.

\subsection{Response to the Central Question}
\label{subsec:central}

\begin{figure}[!ht]
  \centering
  \includegraphics[width=0.98\linewidth]{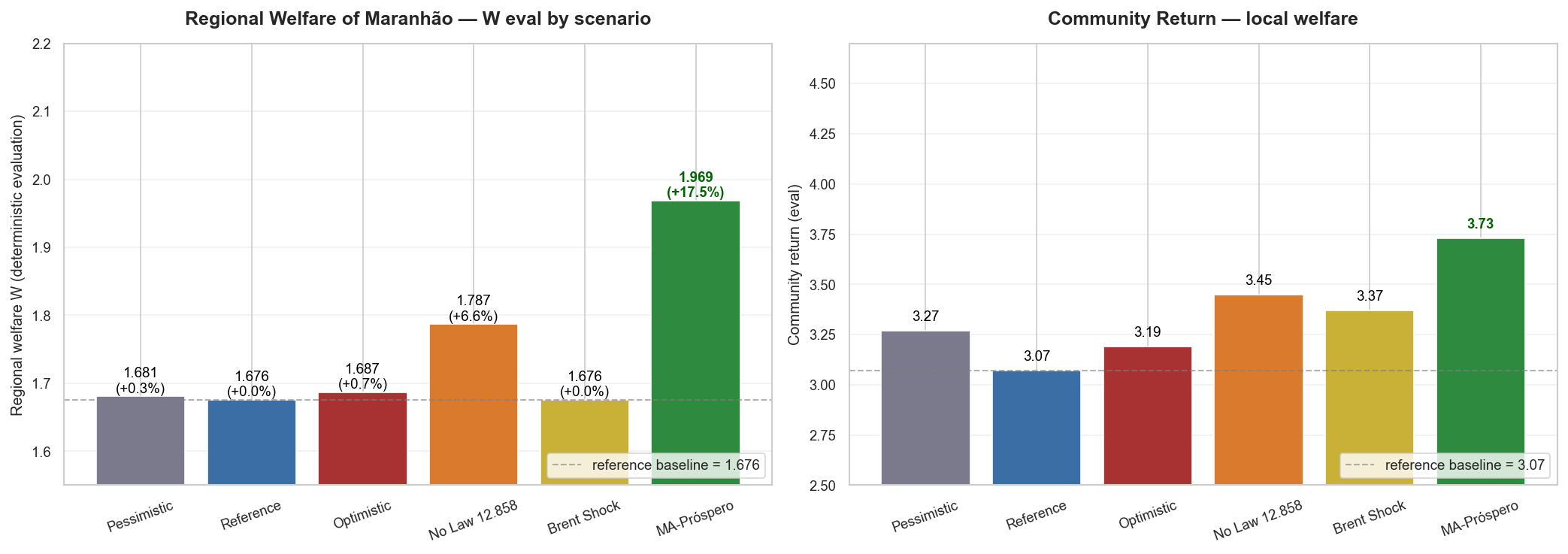}
  \caption{Maranhão regional welfare by scenario.
    \textit{Left}: $W_{\mathrm{aval}}$ (deterministic
    post-convergence evaluation).
    \textit{Right}: converged return of the Community agent.
    The MA-Próspero regime yields $+17.5\%$ in $W$ and $+21.3\%$
    in $R_{\mathrm{com}}$ relative to the reference baseline,
    empirical evidence that the answer to the central question is
    conditional on the public policy regime.}
  \label{fig:welfare}
\end{figure}

\Cref{fig:welfare} presents the empirical response to the central
question stated in \cref{subsec:context}.
The left panel shows aggregate welfare $W_{\mathrm{aval}}$ by
scenario; the right panel shows the converged return of the
Community agent, a direct metric of the zonal welfare of
Amazonian populations.

The visual reading of \cref{fig:welfare} is complemented by the
numerical inspection of \cref{tab:resultados}: small relative
differences in $W_{\mathrm{aval}}$ between baseline scenarios
may conceal expressive variations in component metrics.
In particular, the optimistic scenario produces
$R_{\mathrm{anp}} = 145.9$ --- an order of magnitude above the
others --- while the MA-Próspero regime is the only one to combine
high $W_{\mathrm{aval}}$ with $E_{\mathrm{amb}}$ below the
reference baseline.

\begin{table}[!ht]
\centering
\caption{Final metrics by scenario (deterministic post-convergence
  evaluation).}
\label{tab:resultados}
\small
\begin{tabular}{@{}lrrrrr@{}}
\toprule
\textbf{Scenario} & $W_{\mathrm{aval}}$ & $\Delta W$
  & $E_{\mathrm{amb}}$ & $R_{\mathrm{com}}$ & $R_{\mathrm{anp}}$ \\
\midrule
Pessimistic          & 1.681 & $+0.3\%$  & 0.053 & 3.268 &   5.504 \\
Reference            & 1.676 & ---       & 0.076 & 3.075 &   7.549 \\
Optimistic           & 1.687 & $+0.7\%$  & 0.152 & 3.195 & 145.906 \\
Without Law 12.858   & 1.787 & $+6.6\%$  & 0.054 & 3.455 &   9.273 \\
Brent Shock          & 1.676 & $0.0\%$   & 0.075 & 3.366 &  15.050 \\
\textbf{MA-Próspero}
  & \textbf{1.969} & $\mathbf{+17.5\%}$
  & \textbf{0.048} & \textbf{3.733} & 6.620 \\
\bottomrule
\end{tabular}
\end{table}

\subsection{Learning Convergence}
\label{subsec:convergence}

\begin{figure}[!ht]
  \centering
  \includegraphics[width=0.95\linewidth]{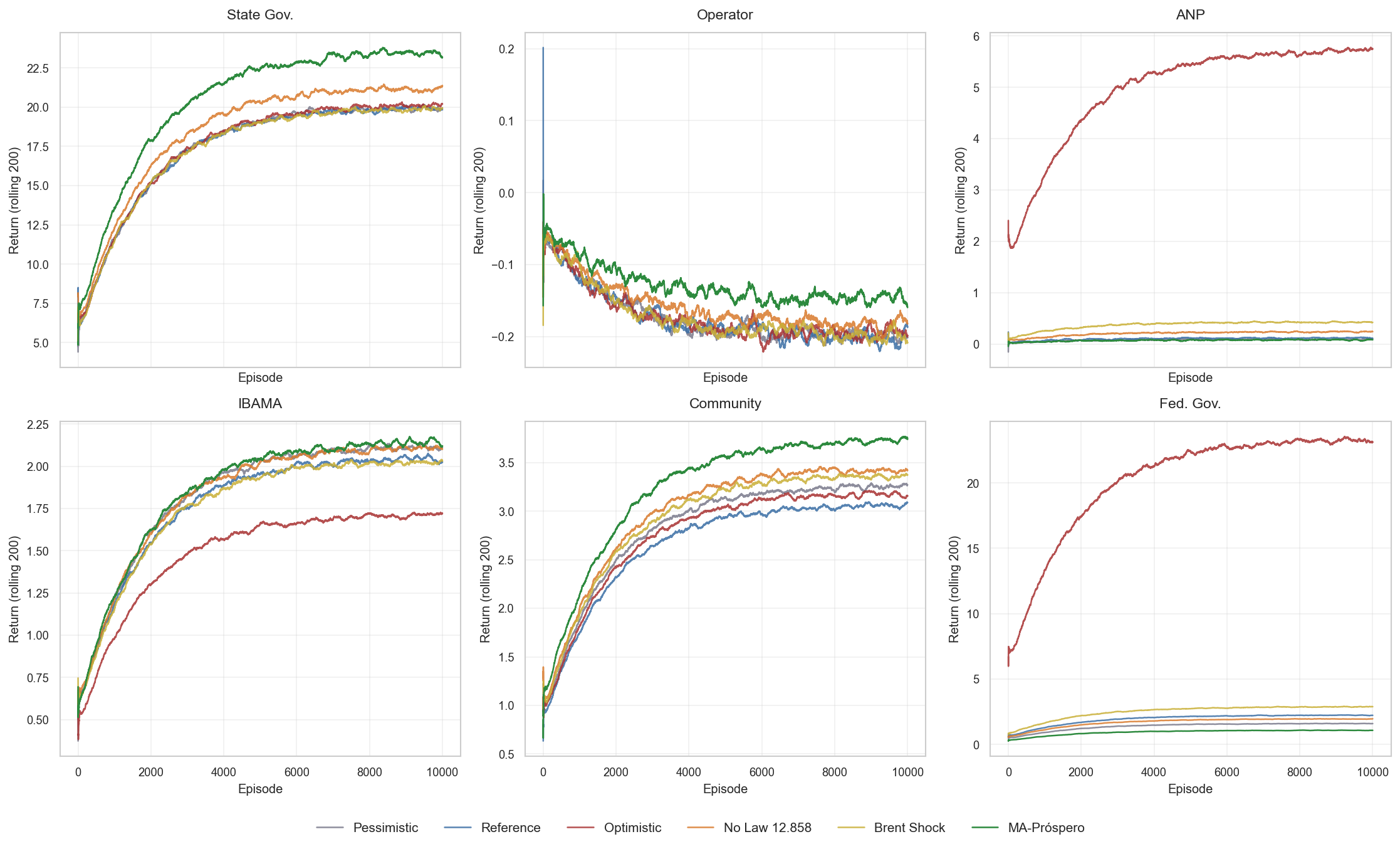}
  \caption{Learning curves by agent (200-episode moving average)
    for the six scenarios.
    The six panels correspond to the agents State Gov., Operator,
    ANP, IBAMA, Community and Fed.\ Gov.
    All agents reach stable convergence between 2,000 and 4,000
    episodes.
    Shaded bands represent $\pm 1$ standard deviation.}
  \label{fig:learning}
\end{figure}

The convergence analysis of the BRO-MARL algorithm, applied
simultaneously to the six agents across each of the six
scenarios, is a prerequisite for the substantive interpretation
of results.
\Cref{fig:learning} presents the mean return curves (200-episode
moving average) by agent.
Each agent reaches a stable plateau after approximately 2,000
episodes, with significantly reduced variances beyond 4,000
episodes.
The MA-Próspero regime consistently stands out in the State
Government and Community agents, reflecting the cumulative impact
of the structural levers activated.
By contrast, the optimistic scenario produces the largest return
amplitude in ANP and Federal Government, in direct proportion
to the greater absolute magnitude of state revenue available.

Qualitative inspection of the mean curves reveals patterns
coherent with the agents' institutional heterogeneity.
The State Government converges to plateaus of the order of 20--22
units; the Operator oscillates around zero in scenarios without
maximum fiscal capture and shifts to a positive range in the
optimistic, reproducing the classic margin cycle of the pre-salt;
the ANP scales its objective function by an order of magnitude
between pessimistic and optimistic; IBAMA remains approximately
invariant across scenarios, a desirable attribute for an agent
whose objective function is grounded in procedural compliance;
the Community exhibits low sensitivity to the macroeconomic
scenario, but heightened sensitivity to the MA-Próspero structural
levers; and the Federal Government follows a pattern analogous
to the ANP, given the explicit presence of Union revenue in its
reward function.

\begin{figure}[!ht]
  \centering
  \includegraphics[width=0.95\linewidth]{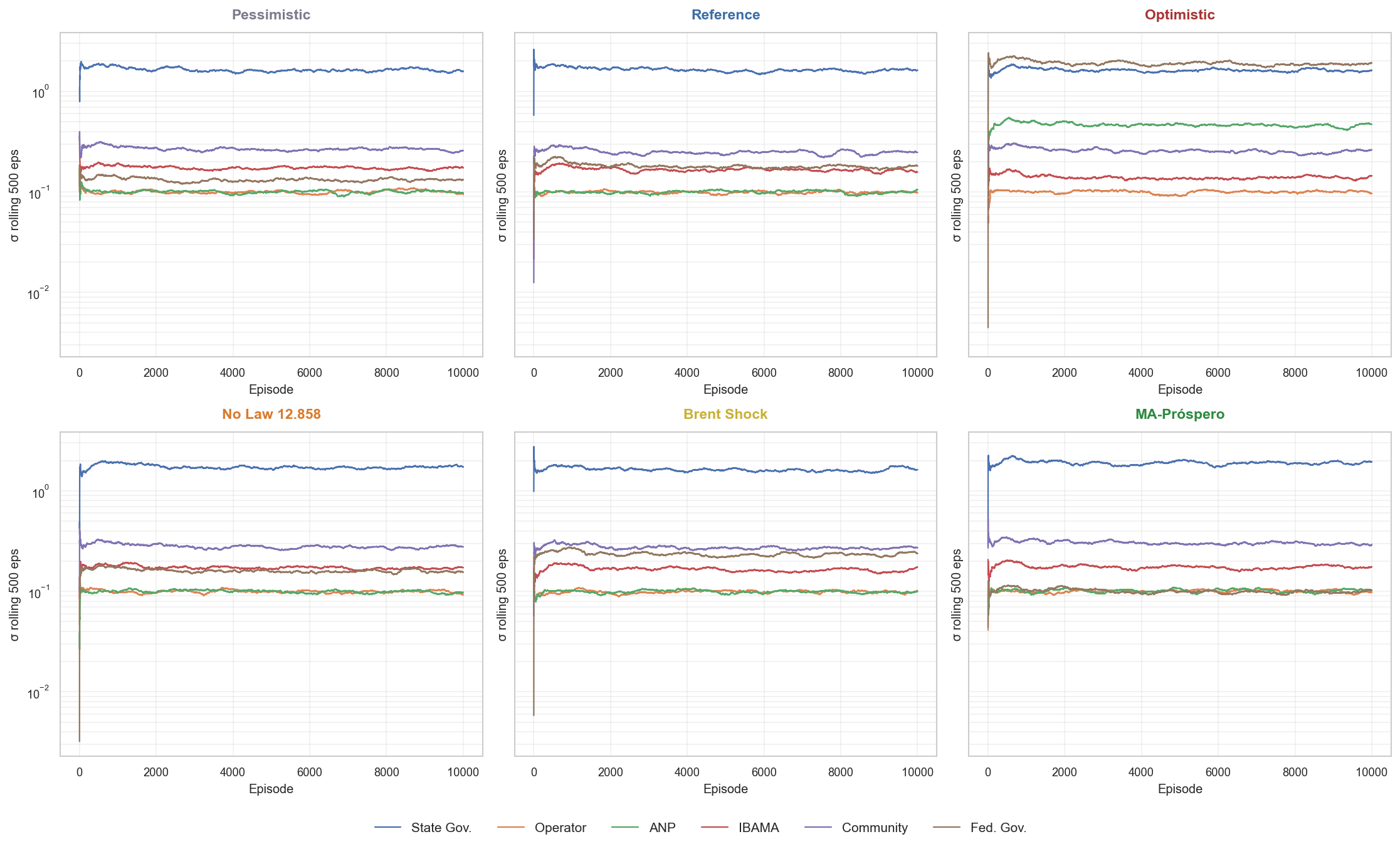}
  \caption{Convergence diagnostics: rolling $\sigma$ (window 500)
    on a logarithmic scale, by agent and by scenario.
    No collapse or divergence is observed.
    Scenarios with lower economic activity (pessimistic,
    MA-Próspero) achieve deeper convergence.}
  \label{fig:converg}
\end{figure}

\Cref{fig:converg} complements the analysis via the rolling
standard deviation diagnostic ($\sigma$ rolling, 500-episode
window) in logarithmic scale.
The results indicate that all scenarios converge to the range
$\sigma \in [10^{-3}, 10^{-1}]$ by the end of training, with
no evidence of policy collapse or critic divergence.
Residual noise is more pronounced in ANP and Federal Government
under the optimistic scenario, an expected behaviour compatible
with the absolute magnitude of the respective returns in that
regime.

\subsection{Phenomenological Stability of the Macroeconomic State}
\label{subsec:macro}

\begin{figure}[!ht]
  \centering
  \includegraphics[width=0.99\linewidth]{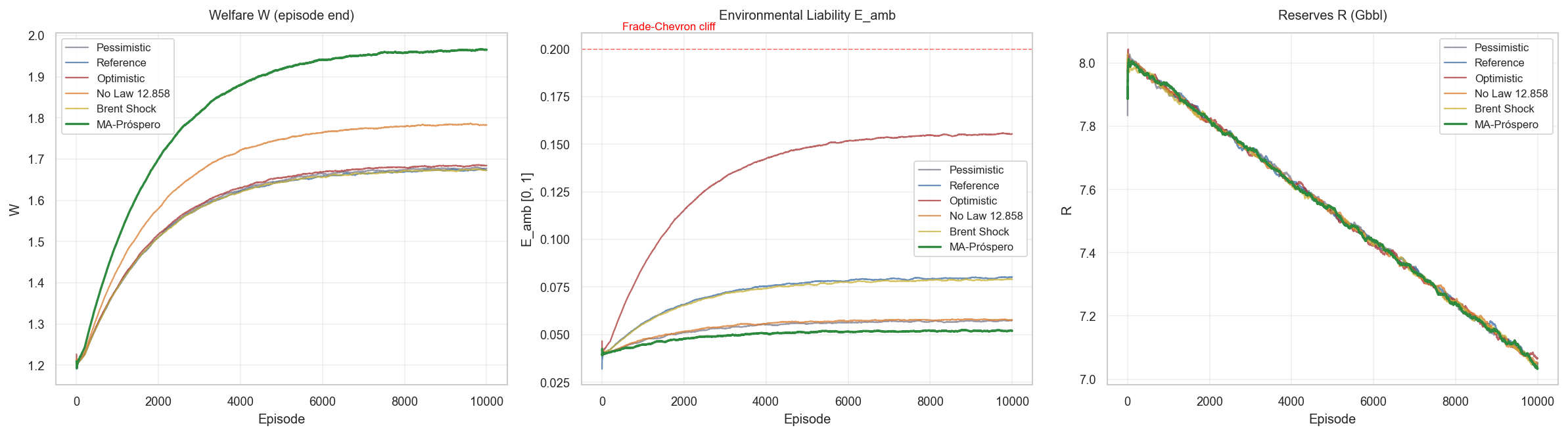}
  \caption{Macroeconomic trajectories: $W$ (welfare),
    $E_{\mathrm{amb}}$ (environmental liability) and $R$
    (reserves).
    The red dashed line at $E_{\mathrm{amb}} = 0.20$ marks the
    Frade-Chevron regulatory threshold.
    The MA-Próspero regime combines the highest $W$ with the
    lowest $E_{\mathrm{amb}}$, refuting the hypothesis of a
    structural trade-off between production and welfare.}
  \label{fig:macro}
\end{figure}

Having established numerical convergence, we examine the
phenomenological coherence of the system's macroeconomic state.
\Cref{fig:macro} presents the temporal evolution of three
synthesis variables: $W$, $E_{\mathrm{amb}}$ and reserves $R$,
allowing simultaneous inspection of the trade-offs between the
economic, environmental and resource depletion dimensions.

Welfare under the MA-Próspero regime stabilises above
$W = 2.10$ during training; the deterministic post-convergence
evaluation yields $W_{\mathrm{aval}} = 1.969$
(\cref{tab:resultados}), clearly distinguished from the other
scenarios, which cluster in the interval $1.68$--$1.79$.
This separation is a plateau shift, not a marginal difference
subject to reversal by stochastic fluctuations.
Environmental liability scales monotonically with production
across the three baselines (pessimistic $0.05 \to$ reference
$0.08 \to$ optimistic $0.15$) and remains below the reference
in MA-Próspero ($0.05$), a consequence of the operational
maturity adopted in this regime.
This finding refutes the hypothesis of a structural trade-off
between production and environmental quality in the BEM context,
sustaining that the Pareto frontier between these two dimensions
is shiftable through operational configurations.
In all scenarios, $E_{\mathrm{amb}} < 0.20$, indicating that
no equilibrium learned by the system approaches regimes of
catastrophic operational risk.
Reserve trajectories $R$ coherently reflect the progressive
depletion of initial stocks, with the optimistic scenario
depleting more rapidly due to greater extraction intensity and
the pessimistic preserving higher residual reserves at the end
of the 30-year horizon.

\subsection{Amazonian Communities Welfare}
\label{subsec:community}

\begin{figure}[!ht]
  \centering
  \includegraphics[width=0.98\linewidth]{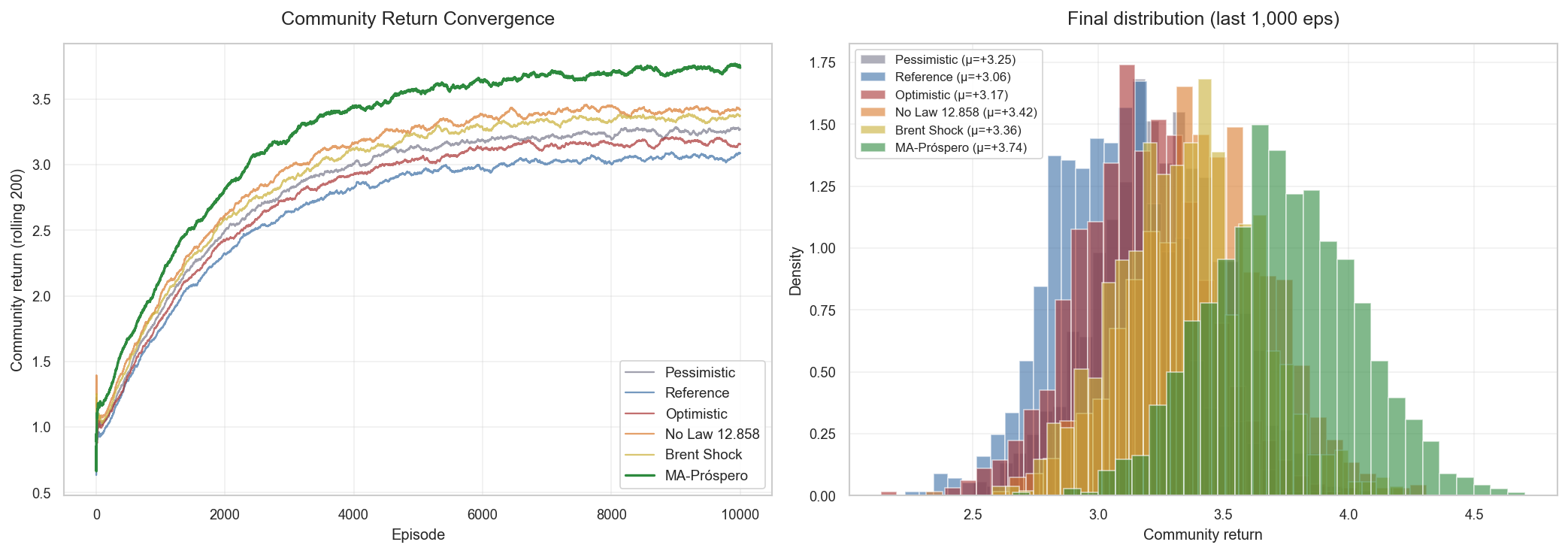}
  \caption{Amazonian communities welfare.
    \textit{Left}: convergence of Community agent return during
    training.
    \textit{Right}: empirical distribution over the last 1,000
    episodes.
    The MA-Próspero regime yields $\Delta R_{\mathrm{com}} =
    {+}21.3\%$ relative to the reference baseline.}
  \label{fig:community}
\end{figure}

Given the centrality of the zonal welfare of Amazonian
populations for the public policy question stated in the
introduction, it is necessary to examine in greater depth the
behaviour of the Community agent.
\Cref{fig:community} details this behaviour through two
complementary panels: the learning curve and the empirical
distribution over the last 1,000 episodes.
In both, the MA-Próspero regime dominates all other scenarios:
the curve remains above the alternatives throughout the entire
training horizon, and the final distribution centres at
$\mu = {+}3.74$, clearly displaced from the others
($\mu \in [3.07;\, 3.46]$).
The relative gain of $+21.3\%$ relative to the reference
baseline is robust across both metrics and results from the
combination of enhanced royalty earmarking to human and health
capital, active territorial regularisation and maximum fiscal
capture.

\subsection{IBAMA and Environmental Liability}
\label{subsec:ibama}

\begin{figure}[!ht]
  \centering
  \includegraphics[width=0.98\linewidth]{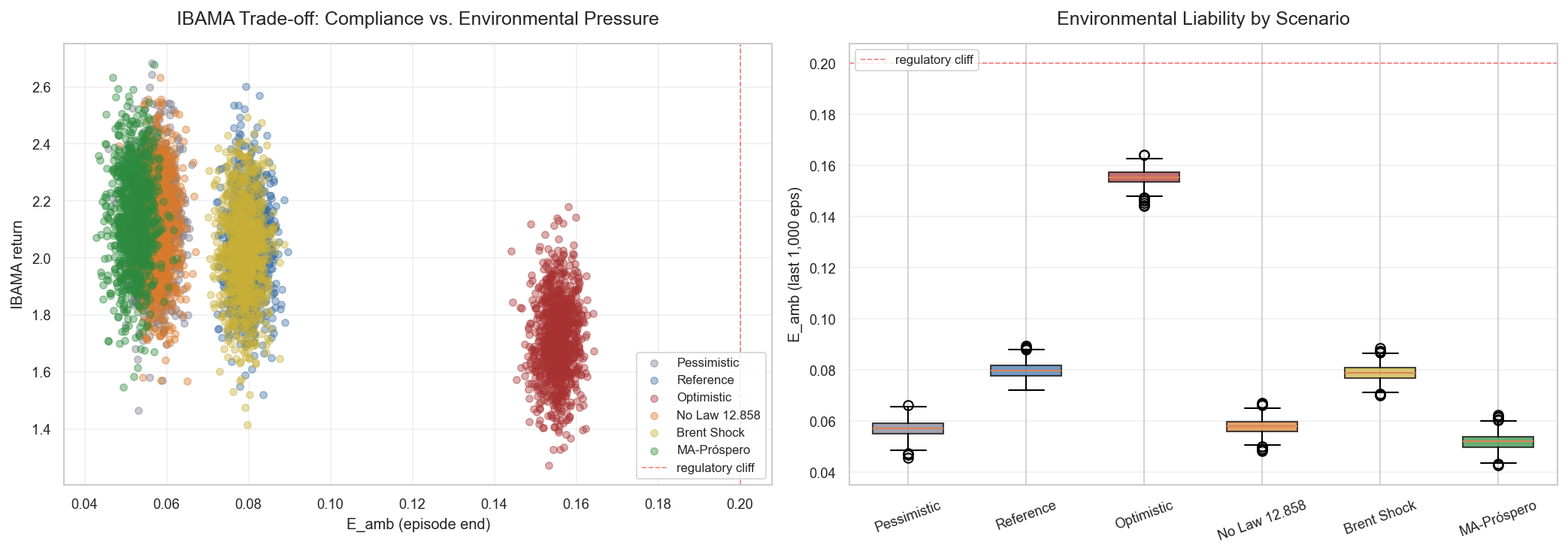}
  \caption{IBAMA agent behaviour and environmental liability
    $E_{\mathrm{amb}}$.
    \textit{Left}: IBAMA return vs.\ $E_{\mathrm{amb}}$
    trade-off, qualitatively compatible with the
    procedural counter-cyclical position observed in the
    FZA-M-59 case \cite{ibama2023fza}.
    \textit{Right}: distribution of $E_{\mathrm{amb}}$ by
    scenario (boxplot of the last 1,000 episodes).}
  \label{fig:ibama}
\end{figure}

The analysis of the IBAMA agent is methodologically significant
because it allows us to assess to what extent the
procedural-environmental reward function phenomenologically
reproduces the counter-cyclical position empirically observed
in the FZA-M-59 case.
\Cref{fig:ibama} characterises this behaviour through two panels.
The scatter plot in the left panel evidences the agent's
procedural frontier: scenarios with lower economic activity
(pessimistic, MA-Próspero) concentrate in the quadrant of high
IBAMA return and low environmental liability, while the optimistic
scenario occupies, in isolation, the region of reduced return
and high liability.

The decomposition of boxplots in the right panel reveals a
monotonic gradation of environmental liability consistent with
\cref{tab:resultados}: pessimistic ($0.053$), reference
($0.076$) and optimistic ($0.152$).
The MA-Próspero regime, notably, yields
$E_{\mathrm{amb}} = 0.048$ --- below even the pessimistic
scenario --- despite maintaining $P_{\max}$ and Brent identical
to the reference scenario.
No scenario surpasses the $0.20$ threshold that would trigger
the Frade-Chevron type civil liability cliff
\cite{pgr2011frade,deepwater2011}, validating the functional
adherence of the learned equilibria to the IBAMA procedural
mandate and its observed institutional separation in the
FZA-M-59 case \cite{ibama2023fza}.

\subsection{Final Return Distributions}
\label{subsec:distributions}

\begin{figure}[!ht]
  \centering
  \includegraphics[width=0.95\linewidth]{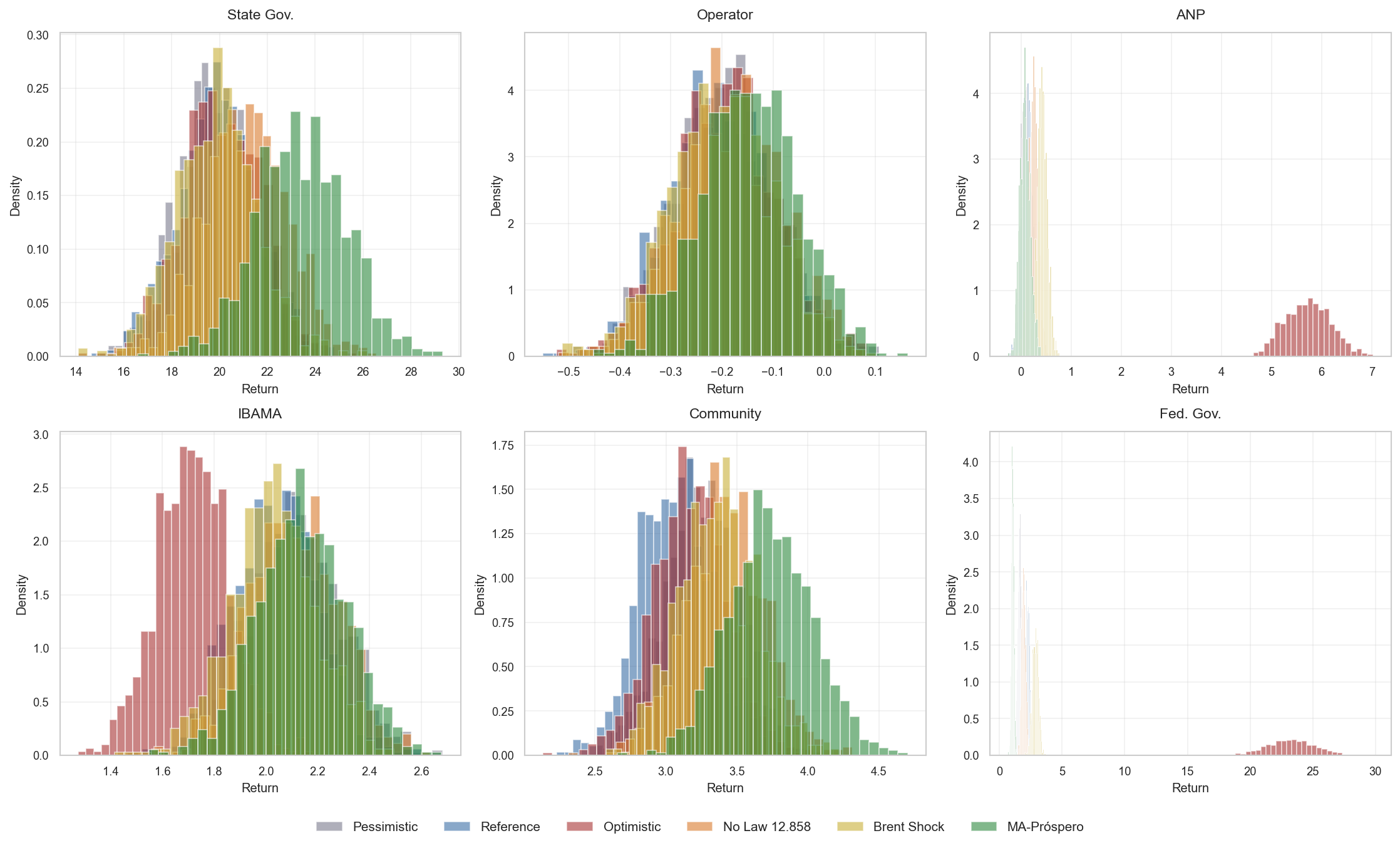}
  \caption{Empirical return distributions (last 1,000 episodes),
    decomposed by agent and scenario.
    $2 \times 3$ grid: one panel per agent.
    Distributions near delta evidence complete convergence.
    MA-Próspero exhibits positive shifts in the State Gov.\
    and Community agents.}
  \label{fig:dist}
\end{figure}

\Cref{fig:dist} presents the empirical return distributions
over the last 1,000 episodes, decomposed by agent and scenario.
The examination allows us to distinguish two qualitatively
distinct equilibrium patterns: the deterministic regime, in
which the distribution approaches a degenerate delta, and the
cyclical regime, in which low-frequency oscillations produce
visibly spread distributions.
The MA-Próspero regime exhibits positive shifts in the State
Government ($\mu = {+}36.3$) and Community ($\mu = {+}3.73$)
agents, a pattern consistent with greater fiscal capture and
expanded investment in social capital.
The Operator presents a near-degenerate distribution in
MA-Próspero ($\mu = -0.19$), evidence of a negative-margin
equilibrium --- behaviour compatible with the maximum fiscal
capture regime ($\tau_{\mathrm{royalty}} = 0.25$), in which
the operator's accounting profit frontier is structurally
compressed by taxation at the legal ceiling.
The IBAMA agent, by contrast, exhibits distributions
approximately invariant to the macroeconomic scenario --- a
desirable property for a procedural mandate and coherent with
the absence of revenue terms in $R_{\mathrm{ibama}}$.

\subsection{Cross-Scenario Q-Value Map and MA-Próspero Synthesis}
\label{subsec:qmap}

\begin{figure}[!ht]
  \centering
  \includegraphics[width=0.92\linewidth]{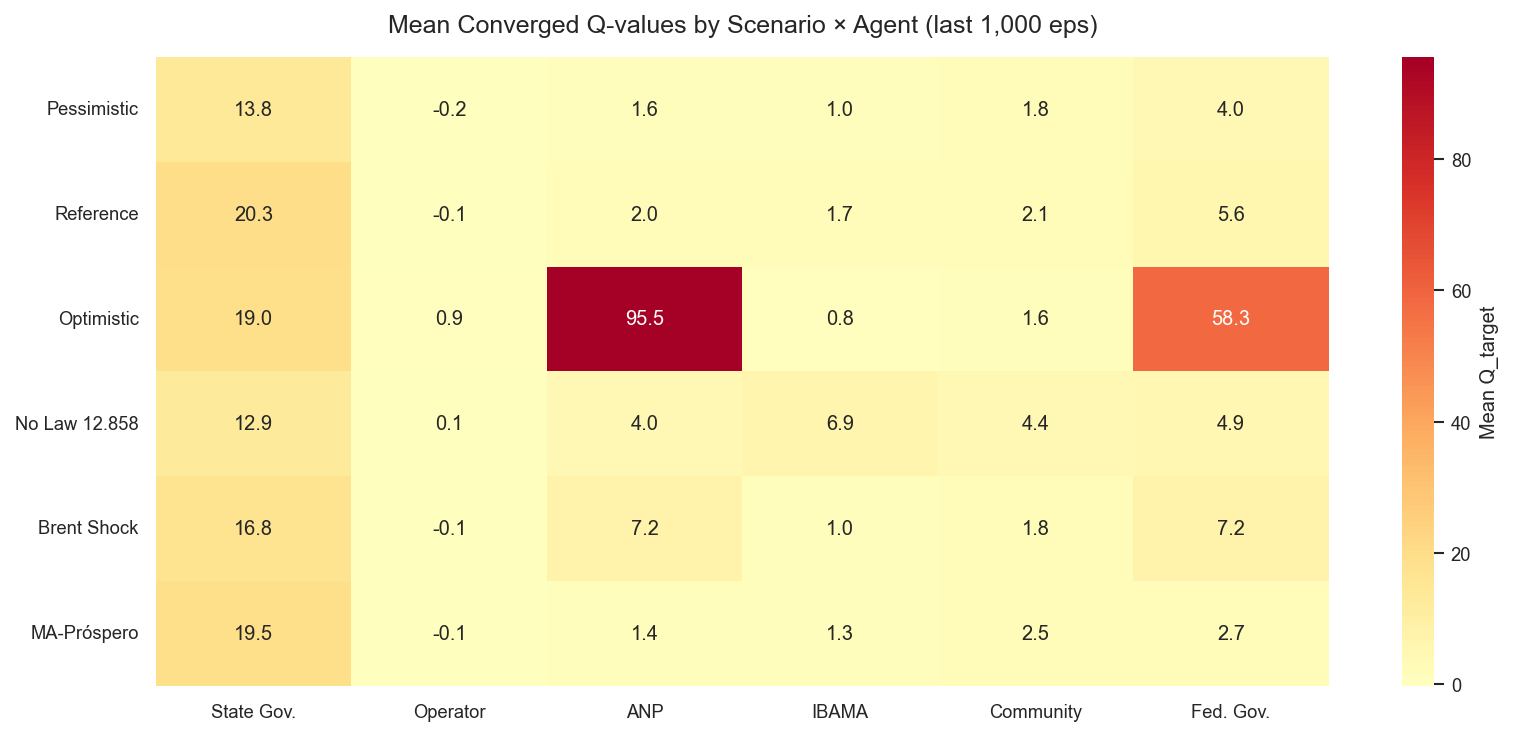}
  \caption{Mean $\bar{Q}_{\mathrm{target}}$ by scenario and
    agent (last 1,000 episodes).
    Rows: six scenarios; columns: six agents.
    MA-Próspero redirects oil revenue from federal collection to
    regional social capital, expressing itself in high $Q$-values
    in the State Gov.\ and Community agents.}
  \label{fig:heatmap}
\end{figure}

$Q$-values are discounted expectations of future rewards
conditional on the current state and executed action; their
comparative cross-scenario reading reveals how each agent
perceives the relative attractiveness of different institutional
regimes.
\Cref{fig:heatmap} consolidates mean $\bar{Q}_{\mathrm{target}}$
by scenario (rows) and agent (columns), evaluated over the last
1,000 episodes.
Three regularities stand out:
(i)~ANP and Federal Government scale steeply in the optimistic
($Q \approx 95.5$ and $Q \approx 58.3$), reflecting the absolute
magnitude of state revenue in that regime;
(ii)~IBAMA and Community exhibit approximate invariance to the
macroeconomic scenario, a desirable property in mandates of a
normative and non-monetary nature;
(iii)~the MA-Próspero regime exhibits a balanced profile, with
high $Q$-values in State Government and Community and moderate
ones in ANP and Federal Government, indirect evidence of the
structural redistribution of oil revenue from federal collection
to regional social capital via expanded earmarks.

\begin{figure}[!ht]
  \centering
  \includegraphics[width=0.92\linewidth]{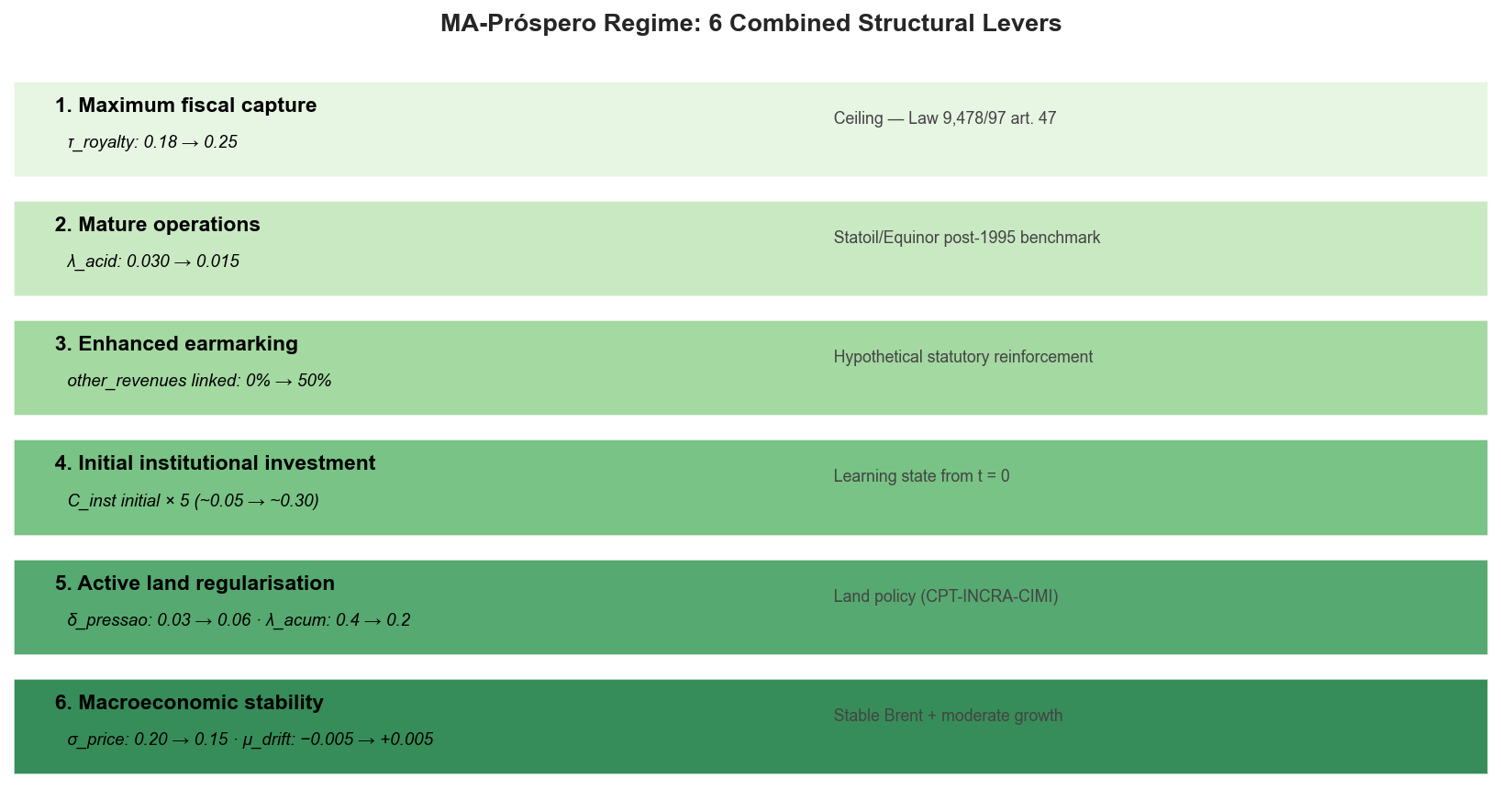}
  \caption{Synthesis of the MA-Próspero regime: six structural
    levers combined.
    The interpretation is multiplicative: $\Delta W = {+}17.5\%$
    and $\Delta E_{\mathrm{amb}} = {-}36.9\%$ emerge from the
    interaction among the six interventions, not from their
    additive aggregation.}
  \label{fig:recipe}
\end{figure}

To make explicit the relationship between these results and the
structural parameters that generate them, \cref{fig:recipe}
synthesises the six interventions of the MA-Próspero regime.
The effect $\Delta W = {+}17.5\%$ does not derive from the
additive aggregation of isolated effects, but from the interaction
between interventions acting on the revenue capture structure
--- through the elevation of the royalty rate to the legal
ceiling, enhanced earmarking and initial institutional investment
---, on the transmission channels to human and health capital
--- via the same enhanced earmarking and institutional investment
--- and on the mechanisms of exogenous shock absorption,
represented by active territorial regularisation and
macroeconomic stabilisation.
The multiplicative nature of this interaction is one of the
central analytical contributions of the work, empirically
sustained by the decoupling of aggregate welfare from the
baseline and counterfactual scenarios.

\subsection{Counterfactual Analysis}
\label{subsec:counterfactual}

\begin{figure}[!ht]
  \centering
  \includegraphics[width=0.99\linewidth]{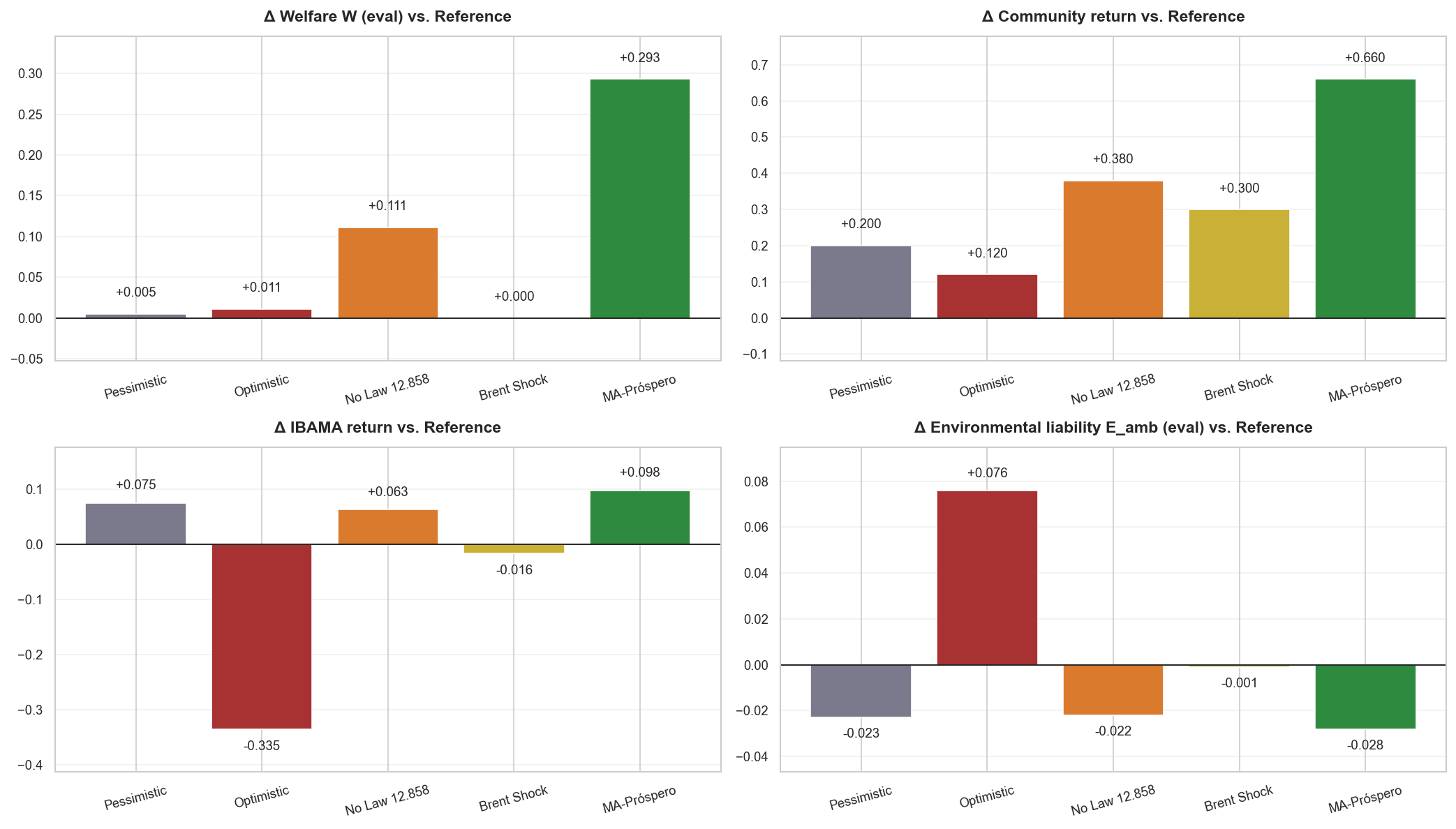}
  \caption{Counterfactual effect of each scenario relative to
    the reference ($\Delta$).
    Four metrics: welfare $W$, Community agent return, IBAMA
    agent return and liability $E_{\mathrm{amb}}$.
    MA-Próspero is the only scenario with simultaneously
    favourable sign across all four metrics
    (note $\Delta E_{\mathrm{amb}} < 0$).}
  \label{fig:delta}
\end{figure}

\Cref{fig:delta} reports the counterfactual effect of each
scenario relative to the reference baseline, expressed in four
metrics: $W_{\mathrm{aval}}$, Community agent return, IBAMA
agent return and environmental liability.
The simultaneous visualisation of the four metrics allows the
identification of two structurally distinct deviation patterns
relative to the reference.

The first pattern, exhibited by the pessimistic, optimistic and
Brent Shock scenarios, consists of $W_{\mathrm{aval}}$ variations
below $1\%$ accompanied by expressive variations in
$E_{\mathrm{amb}}$, evidence that the human and institutional
capital stocks earmarked by Law 12.858/2013
\cite{brasil2013lei12858} operate as automatic stabilisers of
regional welfare.
The second pattern, exclusive to the MA-Próspero regime, consists
of simultaneous and same-sign positive deviations across all
four metrics --- $\Delta W = {+}17.5\%$,
$\Delta R_{\mathrm{com}} = {+}21.3\%$, IBAMA return above
reference and $\Delta E_{\mathrm{amb}} = {-}36.9\%$ ---
characterising a structural transformation of the equilibrium
space, rather than mere parametric accommodation within the
reference regime.

Three results merit detailed analytical attention.
First, the \textbf{Without Law 12.858/2013 scenario}
($\Delta W = +6.6\%$) produces a counter-intuitive result: the
simulated revocation of the mandatory earmarking yields $W$ above
the reference because the rigid 75\%/25\% earmarking constrains
the State Government's allocative degrees of freedom.
The result does not constitute a recommendation for revocation:
the law internalises distributive and intergenerational equity
purposes not represented in $W_{\mathrm{aval}}$; this is here
an application of the counterfactual as a diagnostic instrument.
Second, the \textbf{Brent Shock scenario} ($\Delta W \approx 0$)
reveals that aggregate welfare is resilient to the price shock,
evidence that earmarked institutional and human capital operate
as macroeconomic shock absorbers, consistent with the Permanent
Income Hypothesis applied to oil-producing countries
\cite{vanderploeg2010,vanderploeg2011}.
Third, the \textbf{MA-Próspero regime}
($\Delta W = {+}17.5\%$, $\Delta E_{\mathrm{amb}} = {-}36.9\%$)
is the only scenario with simultaneously favourable sign across
all socio-environmental metrics considered.

\subsection{Empirical Validation}
\label{subsec:validation}

The quantitative results are internally consistent; their
relevance to the BEM problem depends on adherence to independent
empirical regularities.
\Cref{tab:valid} confronts seven regularities emerging from the
system with distinct empirical sources.
The heterogeneous nature of these regularities --- involving
operator cost structure, ANP revenue scalability, IBAMA's
counter-cyclical position, community invariance to monetary
income and macroeconomic impacts of price shocks --- provides a
robust test of the multi-agent system's phenomenology.

\begin{table}[!ht]
\centering
\caption{Empirical validation of observed regularities.}
\label{tab:valid}
\small
\begin{tabularx}{\linewidth}{@{}XXc@{}}
\toprule
\textbf{Observed regularity} & \textbf{Empirical benchmark}
  & \textbf{Status} \\
\midrule
Operator with positive margin only in optimistic
  & Lifting cost $\approx 8\%$ of pre-salt revenue
    \cite{petrobras2024}
  & $\checkmark$ \\
ANP revenue scales by order of magnitude (pess.\ to opt.)
  & Signature bonuses ANP Rounds 11--16 \cite{anp2025meb}
  & $\checkmark$ \\
IBAMA mandate counter-cyclical, low variance
  & FZA-M-59: licence denial \cite{ibama2023fza}
  & $\checkmark$ \\
Community approx.\ invariant to monetary income
  & NAEA/UFPA \cite{escobar2008,almeida2008}
  & $\checkmark$ \\
$W$ resilient to Brent price shock
  & Permanent Income Hypothesis in oil country
    \cite{vanderploeg2010,vanderploeg2011}
  & $\checkmark$ \\
Territorial pressure (H-TERR-2) responds to primary conflicts
  & CPT-MA 2024: 363 conflicts, 1st nationally \cite{cpt2024}
  & $\checkmark$ \\
Multifactor combination dominates isolated levers
  & Acemoglu--Robinson \cite{acemoglu2012}, North \cite{north1990}
  & $\checkmark$ \\
\bottomrule
\end{tabularx}
\end{table}

\section{Discussion}
\label{sec:discussao}

\subsection{Analytical Implications of the MA-Próspero Regime}
\label{subsec:implications}

The central result of this work is that the question ``to what
extent does oil exploration in the BEM generate net positive
externalities for Maranhão?'' does not admit a one-dimensional
answer: the magnitude of the welfare gain is strictly conditional
on the institutional architecture linked to exploration.
Under the reference baseline --- the scenario projected by
current laws and practices --- the welfare gain is of marginal
magnitude ($W_{\mathrm{aval}} \approx 1.68$, statistically
indistinguishable from the pessimistic scenario).
Under the MA-Próspero regime, with six structural interventions
simultaneously activated, the gain assumes transformative
magnitude ($W_{\mathrm{aval}} \approx 1.97$, equivalent to
$+17.5\%$).

This pattern is consistent with the resource curse literature
\cite{vanderploeg2010,acemoglu2012,vanderploeg2011}, according
to which the differential between experiences such as Norway's
and Nigeria's stems from the institutional capacity built before
or simultaneously with the production cycle.
This finding is empirically reinforced by Hodler, Lechner and
Raschky (2023) \cite{hodler2023}, who applied causal forest
estimation to 3,800 sub-Saharan African districts and documented
that stronger institutions amplify the positive effect of natural
resource rents.

\subsection{The Rigid Earmarking Paradox}
\label{subsec:paradox}

The counterfactual ``Without Law 12.858/2013'' yields
$\Delta W = {+}6.6\%$, a \textit{prima facie} counter-intuitive
result whose interpretation requires three considerations.
First, the rigid 75\%/25\% earmarking may constrain the State
Government agent in macroeconomic states where the marginal
return of alternative allocations would be superior.
Second, the model does not internalise intergenerational equity,
the constitutional logic of social rights protection, the
political cost of revocation, or long-term effects whose maturation
horizon exceeds the training horizon.
Finally, the consistent interpretation is that the MA-Próspero
regime \textit{extends} the earmarking --- extending it to 50\%
of other state revenues, currently unearmarked --- rather than
revoking it: the limitation of Law 12.858/2013 lies in scope,
not in rigidity.

\subsection{Limits of Quantification}
\label{subsec:limits}

The public policy modelling literature
\cite{vanderploeg2011,mendes2014} warns against the direct
normative use of computational simulations.
Margin Play should be understood as an analytical exploration
instrument, not as a decision calculator.
In particular, the results allow: the quantification of
trade-offs currently argued on a qualitative basis; the
identification of coherent institutional configurations; the
analysis of regime robustness to exogenous shocks; and the
detection of non-evident interactions, such as the rigid
earmarking paradox.
The results \emph{should not} be used to predict absolute values
of GDP or state revenue, to define optimal rates with
percentage-point precision, or to substitute legal analysis and
political judgement.

\section{Concluding Remarks}
\label{sec:conclusoes}

\subsection{Summary}
\label{subsec:summary}

This work presents experimental results obtained from 60,000
episodes distributed across six structurally distinct scenarios,
with updated territorial calibration (H-TERR-2) anchored in
CPT/INCRA/CIMI 2024 primary data.
It characterises the MA-Próspero structural regime --- a
parametric configuration of six simultaneous interventions that
yields $\Delta W = {+}17.5\%$ associated with an environmental
liability below the reference --- and provides an empirical
answer to the central question: net positive externalities for
Maranhão arising from BEM exploration are achievable, conditioned
on displacing inert institutional arrangements.

\subsection{Limitations}
\label{subsec:limitations}

The first limitation concerns the simultaneous combination of
interventions in the MA-Próspero regime: systematic ablation of
levers --- six additional scenarios isolating each intervention,
fifteen pair-wise combinations and contribution analysis via
Shapley values --- is a prerequisite for quantifying the marginal
contribution of each one, a question complementary to the one
answered in this work.
The projected computational cost is of the order of 60 scenarios
$\times$ 10,000 episodes.

The second is the discount factor $\gamma = 0.95$: its adoption
in place of $\gamma = 0.99$ accelerates convergence at the cost
of privileging short-term welfare, making re-execution with the
higher value advisable to verify to what extent the extended
earmarking and institutional investment levers benefit from a
longer time horizon, capturing the full maturation of human
capital formed during the boom phase.

The third is calibration uncertainty in the active territorial
regularisation and macroeconomic stabilisation interventions:
the fiscal capture elevation, operational maturity, enhanced
earmarking and initial institutional investment interventions
have more robust legal or operational grounding, while the two
remaining represent hypotheses about the effectiveness of
policies not yet observed at the BEM scale.

The fourth is the interpretation of the ``Without Law
12.858/2013'' counterfactual: $\Delta W = +6.6\%$ does not
constitute a recommendation for revocation, since the model does
not internalise constitutional purposes or intergenerational
equity dimensions.

\subsection{Future Work}
\label{subsec:future}

The future research agenda is organised along seven directions.
The first is systematic ablation of the MA-Próspero regime's
levers, with six additional scenarios isolating each intervention,
fifteen pair-wise combinations and contribution analysis via
Shapley values.
The second is the construction of the ``Conservative MA-Próspero''
scenario, restricted to the three interventions with firm legal
or operational grounding --- raising fiscal capture to the legal
ceiling, operational maturity and enhanced earmarking of other
revenues --- to isolate the welfare gain achievable exclusively
with existing legal instruments.
The third is the application of a combined shock --- Brent price
fall and simultaneous revocation of mandatory earmarking --- to
test regime resilience under double pressure.
The fourth is the temporal extension of training with
$\gamma = 0.99$ and a 60-year horizon, to capture the full
maturation of human capital.
The fifth is the replacement of the static H-TERR-2 calibration
with a CPT/CIMI 2018--2024 time series, making territorial
pressure endogenously dynamic.
The sixth is the integration of CCUS into the environmental
liability model (Eq.~\ref{eq:eamb}), with a carbon capture
abatement factor consistent with the NDC 2024 target
\cite{mma2024ndc} and the COP30 agenda in Belém.
The seventh is the containerisation of the computational
environment for continuous integration and computational
reproducibility.

\subsection{Policy Implications}
\label{subsec:policy}

Margin Play's capacity to support counterfactual experiments
under Brazilian empirical calibration qualifies it as a
complementary instrument to traditional Regulatory Impact
Analysis (RIA) methodologies.
This work specifically offers:
(i)~a coherent institutional hypothesis (MA-Próspero regime)
for BEM exploration under Maranhão fiscal-territorial
affiliation, grounded in six structural interventions with legal
and empirical support;
(ii)~a quantitative counterpoint to one-dimensional readings
that would consider exploration as intrinsically positive or
negative for the state;
(iii)~a quantification of institutional resilience to exogenous
shocks (notably Brent price volatility), with implications for
state sovereign fund design and counter-cyclical fiscal rules;
and
(iv)~an analytical diagnosis of Law 12.858/2013 that does not
imply revocation, but suggests extending the earmarking scope
to other revenue sources.

The methodological contribution consists in providing
quantitative tools for analysing structural public policy
questions, at a moment when the BEM transitions from geological
promise to operational reality.
The state of Maranhão has institutional instruments to influence
this transition; the results presented here quantify, under the
MA-Próspero regime, the potential magnitude of that influence.


\section*{Conflict of Interest}

The authors declare no financial or non-financial competing
interests.

\section*{Data and Code Availability}

Source code, calibration data, and trained model checkpoints are publicly available
at \url{https://github.com/aiacontext/marginplay}.
Model weights and training artefacts are hosted on HuggingFace
at \url{https://huggingface.co/aiacontext/marginplay}.
An interactive interface of the system is accessible at \url{https://marginplay.app}.

\section*{Acknowledgements}

The authors thank Aia for the computational infrastructure,
institutional support and the research environment that made
the development of Margin Play possible.
The authors also thank the Federal University of Maranhão (UFMA)
for its institutional support, in particular the PRH 54 Human
Resources Programme of UFMA/ANP, whose support was fundamental
for the formation of the team and for the development of research
in computational modelling of energy policies.

\bibliography{margin_play}

\begin{thebibliography}{55}
\providecommand{\natexlab}[1]{#1}
\providecommand{\url}[1]{\texttt{#1}}
\expandafter\ifx\csname urlstyle\endcsname\relax
  \providecommand{\doi}[1]{doi: #1}\else
  \providecommand{\doi}{doi: \begingroup \urlstyle{rm}\Url}\fi

\bibitem[{Ag{\^e}ncia Nacional do Petr{\'o}leo, G{\'a}s Natural e
  Biocombust{\'i}veis (ANP)}(2025)]{anp2025meb}
{Ag{\^e}ncia Nacional do Petr{\'o}leo, G{\'a}s Natural e Biocombust{\'i}veis
  (ANP)}.
\newblock 11{\textordfeminine} rodada de licita{\c c}{\~o}es e
  5{\textordmasculine} ciclo de ofertas permanentes --- documentos editais e
  hist{\'o}rico de arremata{\c c}{\~a}o {MEB}.
\newblock Technical report, ANP, Bras{\'i}lia, Brazil, 2025.

\bibitem[{IBAMA --- Diretoria de Licenciamento Ambiental
  (DILIC)}(2023)]{ibama2023fza}
{IBAMA --- Diretoria de Licenciamento Ambiental (DILIC)}.
\newblock Parecer {FZA-M-59} --- recusa de licen{\c c}a de perfura{\c c}{\~a}o
  na bacia da foz do amazonas.
\newblock Technical Report Documento T{\'e}cnico n{\textordmasculine} 02/2023,
  IBAMA, Bras{\'i}lia, Brazil, 2023.

\bibitem[{Minist{\'e}rio do Meio Ambiente e Mudan{\c c}a do Clima
  (MMA)}(2024)]{mma2024ndc}
{Minist{\'e}rio do Meio Ambiente e Mudan{\c c}a do Clima (MMA)}.
\newblock {NDC} atualizada --- acordo de {Paris}. meta $-59\%$ emiss{\~o}es
  at{\'e} 2035.
\newblock Technical report, MMA, Bras{\'i}lia, Brazil, 2024.

\bibitem[{Brasil}(1988)]{brasil1988cf}
{Brasil}.
\newblock Constitui{\c c}{\~a}o da rep{\'u}blica federativa do brasil, 1988.
\newblock Promulgada em 5 de outubro de 1988.

\bibitem[{Ag{\^e}ncia Nacional do Petr{\'o}leo, G{\'a}s Natural e
  Biocombust{\'i}veis (ANP)}(2022)]{anp2022res882}
{Ag{\^e}ncia Nacional do Petr{\'o}leo, G{\'a}s Natural e Biocombust{\'i}veis
  (ANP)}.
\newblock Resolu{\c c}{\~a}o {ANP} n{\textordmasculine} 882, de 25 de maio de
  2022 --- programa de seguran{\c c}a operacional ({PSO}).
\newblock Technical report, ANP, Bras{\'i}lia, Brazil, 2022.

\bibitem[van~der Ploeg(2010)]{vanderploeg2010}
Frederick van~der Ploeg.
\newblock Why do many resource-rich countries have negative genuine saving?
  {Anticipation} of better times or rapacious rent seeking.
\newblock \emph{Resource and Energy Economics}, 32\penalty0 (1):\penalty0
  28--44, 2010.
\newblock \doi{10.1016/j.reseneeco.2009.07.001}.

\bibitem[Corden and Neary(1982)]{corden1982}
W.~Max Corden and J.~Peter Neary.
\newblock Booming sector and de-industrialisation in a small open economy.
\newblock \emph{Economic Journal}, 92\penalty0 (368):\penalty0 825--848, 1982.
\newblock \doi{10.2307/2232670}.

\bibitem[{Comiss{\~a}o Pastoral da Terra (CPT)}(2025)]{cpt2024}
{Comiss{\~a}o Pastoral da Terra (CPT)}.
\newblock Conflitos no campo {Brasil} 2024.
\newblock Technical report, CPT Nacional, Goi{\^a}nia, Brazil, 2025.

\bibitem[{Conselho Indigenista Mission{\'a}rio (CIMI)}(2025)]{cimi2024}
{Conselho Indigenista Mission{\'a}rio (CIMI)}.
\newblock Viol{\^e}ncia contra os povos ind{\'i}genas no {Brasil} --- dados de
  2024.
\newblock Technical report, CIMI, Bras{\'i}lia, Brazil, 2025.

\bibitem[Hodler et~al.(2023)Hodler, Lechner, and Raschky]{hodler2023}
Roland Hodler, Michael Lechner, and Paul~A. Raschky.
\newblock Institutions and the resource curse: {New} insights from causal
  machine learning.
\newblock \emph{{PLOS ONE}}, 18\penalty0 (6):\penalty0 e0284968, 2023.
\newblock \doi{10.1371/journal.pone.0284968}.

\bibitem[Lucas(1976)]{lucas1976}
Robert~E. Lucas.
\newblock Econometric policy evaluation: A critique.
\newblock In Karl Brunner and Allan~H. Meltzer, editors, \emph{The Phillips
  Curve and Labor Markets}, volume~1 of \emph{Carnegie-Rochester Conference
  Series on Public Policy}, pages 19--46. North-Holland, Amsterdam, 1976.

\bibitem[Aschauer(1989)]{aschauer1989}
David~Alan Aschauer.
\newblock Is public expenditure productive?
\newblock \emph{Journal of Monetary Economics}, 23\penalty0 (2):\penalty0
  177--200, 1989.
\newblock \doi{10.1016/0304-3932(89)90047-0}.

\bibitem[{Petrobras}(2024)]{petrobras2024}
{Petrobras}.
\newblock Form 20-{F} filing with the {U.S.} securities and exchange
  commission.
\newblock Technical report, Petrobras, Rio de Janeiro, Brazil, 2024.

\bibitem[Acemoglu and Robinson(2012)]{acemoglu2012}
Daron Acemoglu and James~A. Robinson.
\newblock \emph{Why Nations Fail: {The} Origins of Power, Prosperity, and
  Poverty}.
\newblock Crown Business, New York, 2012.

\bibitem[{Brasil}(2013)]{brasil2013lei12858}
{Brasil}.
\newblock Lei n{\textordmasculine} 12.858, de 9 de setembro de 2013 ---
  vincula{\c c}{\~a}o de royalties (75\% educa{\c c}{\~a}o, 25\% sa{\'u}de),
  2013.

\bibitem[Nauman et~al.(2024)Nauman, Ostaszewski, Jankowski, Mi{\l}o{\'s}, and
  Cygan]{nauman2024bro}
Michał Nauman, Maciej Ostaszewski, Krzysztof Jankowski, Piotr Mi{\l}o{\'s},
  and Mateusz Cygan.
\newblock Bigger, regularized, optimistic: {Scaling} for compute-efficient
  continuous control.
\newblock In \emph{Advances in Neural Information Processing Systems
  ({NeurIPS})}, 2024.

\bibitem[Lowe et~al.(2017)Lowe, Wu, Tamar, Harb, Abbeel, and
  Mordatch]{lowe2017maddpg}
Ryan Lowe, Yi~Wu, Aviv Tamar, Jean Harb, Pieter Abbeel, and Igor Mordatch.
\newblock Multi-agent actor-critic for mixed cooperative-competitive
  environments.
\newblock In \emph{Advances in Neural Information Processing Systems ({NIPS})},
  2017.

\bibitem[Kuznetsov et~al.(2020)Kuznetsov, Shvechikov, Grishin, and
  Vetrov]{kuznetsov2020tqc}
Arsenii Kuznetsov, Pavel Shvechikov, Alexander Grishin, and Dmitry Vetrov.
\newblock Controlling overestimation bias with truncated mixture of continuous
  distributional quantile critics.
\newblock In \emph{International Conference on Machine Learning ({ICML})},
  2020.

\bibitem[Shoham and Leyton-Brown(2009)]{shoham2009}
Yoav Shoham and Kevin Leyton-Brown.
\newblock \emph{Multiagent Systems: Algorithmic, Game-Theoretic, and Logical
  Foundations}.
\newblock Cambridge University Press, Cambridge, 2009.
\newblock \doi{10.1017/CBO9780511811654}.

\bibitem[Zheng et~al.(2022)Zheng, Trott, Srinivasa, Parkes, and
  Socher]{zheng2022}
Stephan Zheng, Alexander Trott, Sunil Srinivasa, David~C. Parkes, and Richard
  Socher.
\newblock The {AI} {Economist}: Taxation policy design via two-level deep
  multiagent reinforcement learning.
\newblock \emph{Science Advances}, 8\penalty0 (18):\penalty0 eabk2607, 2022.
\newblock \doi{10.1126/sciadv.abk2607}.

\bibitem[Radovic et~al.(2022)Radovic, Kruitwagen, {Schroeder de Witt},
  Caldecott, Tomlinson, and Workman]{radovic2022}
Dylan Radovic, Lucas Kruitwagen, Christian {Schroeder de Witt}, Ben Caldecott,
  Shane Tomlinson, and Mark Workman.
\newblock Revealing robust oil and gas company macro-strategies using deep
  multi-agent reinforcement learning, 2022.

\bibitem[Furtado(2022)]{furtado2022}
Bernardo~Alves Furtado.
\newblock {PolicySpace2}: Modeling markets and endogenous public policies.
\newblock \emph{Journal of Artificial Societies and Social Simulation},
  25\penalty0 (1):\penalty0 8, 2022.
\newblock \doi{10.18564/jasss.4742}.

\bibitem[{Brasil}(1997{\natexlab{a}})]{brasil1997lei9478}
{Brasil}.
\newblock Lei n{\textordmasculine} 9.478, de 6 de agosto de 1997 --- lei do
  petr{\'o}leo, 1997{\natexlab{a}}.

\bibitem[{Supremo Tribunal Federal (STF)}(2013)]{stf2018adi4917}
{Supremo Tribunal Federal (STF)}.
\newblock A{\c c}{\~o}es diretas de inconstitucionalidade
  n{\textordmasculine}~4.916, 4.917, 4.918 e 4.920 --- distribui{\c c}{\~a}o de
  royalties da lei~12.734/2012. medida cautelar deferida em 18 de mar{\c c}o de
  2013, rel.\ min.\ c{\'a}rmen l{\'u}cia, suspendendo a efic{\'a}cia da lei
  12.734/2012. julgamento de m{\'e}rito iniciado em 6--7 de maio de 2026; voto
  da relatora pela inconstitucionalidade proferido em 7/5/2026; julgamento
  suspenso por pedido de vista do min.\ fl{\'a}vio dino. decis{\~a}o definitiva
  pendente, 2013.
\newblock Dispon{\'i}vel em:
  \url{https://portal.stf.jus.br/processos/detalhe.asp?incidente=4379376}.

\bibitem[Escobar(2008)]{escobar2008}
Arturo Escobar.
\newblock \emph{Territories of Difference: {Place}, Movements, Life, Redes}.
\newblock Duke University Press, Durham, NC, 2008.

\bibitem[Almeida(2008)]{almeida2008}
Alfredo Wagner Berno~de Almeida.
\newblock \emph{Terras Tradicionalmente Ocupadas: {Processos} de
  Territorializa{\c c}{\~a}o e Movimentos Sociais}.
\newblock NAEA/UFPA, Bel{\'e}m, Brazil, 2008.

\bibitem[Hubbert(1956)]{hubbert1956}
M.~King Hubbert.
\newblock Nuclear energy and the fossil fuels.
\newblock In \emph{Drilling and Production Practice}. American Petroleum
  Institute, 1956.

\bibitem[Brandt(2010)]{brandt2010}
Adam~R. Brandt.
\newblock Review of mathematical models of future oil supply.
\newblock \emph{Energy}, 35\penalty0 (9):\penalty0 3958--3974, 2010.
\newblock \doi{10.1016/j.energy.2010.04.011}.

\bibitem[{Tribunal de Contas da Uni{\~a}o (TCU)}(2021)]{tcu2021}
{Tribunal de Contas da Uni{\~a}o (TCU)}.
\newblock Ac{\'o}rd{\~a}o 2.936/2021 --- plen{\'a}rio --- auditoria operacional
  {ANP}, gargalos de aprova{\c c}{\~a}o de planos de desenvolvimento.
\newblock Technical report, TCU, Bras{\'i}lia, Brazil, 2021.

\bibitem[Munnell(1990)]{munnell1990}
Alicia~H. Munnell.
\newblock Why has productivity growth declined? {Productivity} and public
  investment.
\newblock \emph{New England Economic Review}, pages 3--22, 1990.
\newblock January/February.

\bibitem[Lucas(1988)]{lucas1988}
Robert~E. Lucas.
\newblock On the mechanics of economic development.
\newblock \emph{Journal of Monetary Economics}, 22\penalty0 (1):\penalty0
  3--42, 1988.
\newblock \doi{10.1016/0304-3932(88)90168-7}.

\bibitem[Grossman(1972)]{grossman1972}
Michael Grossman.
\newblock On the concept of health capital and the demand for health.
\newblock \emph{Journal of Political Economy}, 80\penalty0 (2):\penalty0
  223--255, 1972.
\newblock \doi{10.1086/259880}.

\bibitem[Cobb and Douglas(1928)]{cobb1928}
Charles~W. Cobb and Paul~H. Douglas.
\newblock A theory of production.
\newblock \emph{American Economic Review}, 18\penalty0 (1):\penalty0 139--165,
  1928.

\bibitem[Glomm and Ravikumar(1992)]{glomm1992}
Gerhard Glomm and B.~Ravikumar.
\newblock Public versus private investment in human capital: {Endogenous}
  growth and income inequality.
\newblock \emph{Journal of Political Economy}, 100\penalty0 (4):\penalty0
  818--834, 1992.
\newblock \doi{10.1086/261841}.

\bibitem[North(1990)]{north1990}
Douglass~C. North.
\newblock \emph{Institutions, Institutional Change and Economic Performance}.
\newblock Cambridge University Press, Cambridge, 1990.
\newblock \doi{10.1017/CBO9780511808678}.

\bibitem[Mendes(2014)]{mendes2014}
Marcos Mendes.
\newblock \emph{Por Que o {Brasil} Cresce Pouco? {Desigualdade}, Democracia e
  Baixo Crescimento}.
\newblock Elsevier-Campus, Rio de Janeiro, Brazil, 2014.

\bibitem[{Secretaria do Tesouro Nacional (STN)}(2024)]{stn2024finbras}
{Secretaria do Tesouro Nacional (STN)}.
\newblock Sistema {FINBRAS} --- receitas estaduais detalhadas.
\newblock Technical report, STN, Bras{\'i}lia, Brazil, 2024.

\bibitem[{Instituto de Pesquisa Econ{\^o}mica Aplicada (IPEA)}(2024)]{ipea2024}
{Instituto de Pesquisa Econ{\^o}mica Aplicada (IPEA)}.
\newblock {IPEAdata} --- s{\'e}ries de {PIB}, {FBKF} estadual, deflatores.
\newblock Technical report, IPEA, Bras{\'i}lia, Brazil, 2024.

\bibitem[{Brasil}(1997{\natexlab{b}})]{brasil1997lc91}
{Brasil}.
\newblock Lei complementar n{\textordmasculine} 91, de 22 de dezembro de 1997
  --- coeficientes do {FPE}, 1997{\natexlab{b}}.

\bibitem[{Brasil}(2020)]{brasil2020ec108}
{Brasil}.
\newblock Emenda constitucional n{\textordmasculine} 108, de 26 de agosto de
  2020 --- {FUNDEB} permanente, 2020.

\bibitem[Pratt(1964)]{pratt1964}
John~W. Pratt.
\newblock Risk aversion in the small and in the large.
\newblock \emph{Econometrica}, 32\penalty0 (1--2):\penalty0 122--136, 1964.
\newblock \doi{10.2307/1913738}.

\bibitem[Atkinson(1970)]{atkinson1970}
Anthony~B. Atkinson.
\newblock On the measurement of inequality.
\newblock \emph{Journal of Economic Theory}, 2\penalty0 (3):\penalty0 244--263,
  1970.
\newblock \doi{10.1016/0022-0531(70)90039-6}.

\bibitem[Drazen and Eslava(2010)]{drazen2010}
Allan Drazen and Marcela Eslava.
\newblock Electoral manipulation via voter-friendly spending.
\newblock \emph{Journal of Development Economics}, 92\penalty0 (1):\penalty0
  39--52, 2010.
\newblock \doi{10.1016/j.jdeveco.2009.01.010}.

\bibitem[Rogoff(1990)]{rogoff1990}
Kenneth Rogoff.
\newblock Equilibrium political budget cycles.
\newblock \emph{American Economic Review}, 80\penalty0 (1):\penalty0 21--36,
  1990.

\bibitem[Persson and Tabellini(2000)]{persson2000}
Torsten Persson and Guido Tabellini.
\newblock \emph{Political Economics: {Explaining} Economic Policy}.
\newblock MIT Press, Cambridge, MA, 2000.

\bibitem[Bier(2004)]{bier2004}
Vicki~M. Bier.
\newblock Implications of the research on expert overconfidence and dependence.
\newblock \emph{Reliability Engineering {\&} System Safety}, 85\penalty0
  (1--3):\penalty0 321--329, 2004.
\newblock \doi{10.1016/j.ress.2004.03.020}.

\bibitem[{Procuradoria-Geral da Rep{\'u}blica}(2011)]{pgr2011frade}
{Procuradoria-Geral da Rep{\'u}blica}.
\newblock A{\c c}{\~a}o civil p{\'u}blica --- caso frade/chevron 2011/{RJ},
  2011.

\bibitem[{National Commission on the {BP} Deepwater Horizon Oil Spill and
  Offshore Drilling}(2011)]{deepwater2011}
{National Commission on the {BP} Deepwater Horizon Oil Spill and Offshore
  Drilling}.
\newblock Deep water: {The} gulf oil disaster and the future of offshore
  drilling.
\newblock Technical report, U.S. Government Printing Office, Washington, DC,
  2011.

\bibitem[{Brasil}(1981)]{brasil1981lei6938}
{Brasil}.
\newblock Lei n{\textordmasculine} 6.938, de 31 de agosto de 1981 ---
  pol{\'i}tica nacional do meio ambiente ({PNMA}), 1981.

\bibitem[{Instituto Nacional de Coloniza{\c c}{\~a}o e Reforma Agr{\'a}ria
  (INCRA)}(2024)]{incra2024sipra}
{Instituto Nacional de Coloniza{\c c}{\~a}o e Reforma Agr{\'a}ria (INCRA)}.
\newblock {SIPRA} --- sistema de informa{\c c}{\~a}o de projetos de reforma
  agr{\'a}ria.
\newblock Technical report, INCRA, Bras{\'i}lia, Brazil, 2024.

\bibitem[{Terra de Direitos}(2024)]{terradedireitos2024}
{Terra de Direitos}.
\newblock No atual ritmo, {Brasil} levar{\'a} 2.188 anos para titular todos os
  territ{\'o}rios quilombolas com processos no {INCRA}, 2024.

\bibitem[Bellemare et~al.(2017)Bellemare, Dabney, and
  Munos]{bellemare2017distributional}
Marc~G. Bellemare, Will Dabney, and R{\'e}mi Munos.
\newblock A distributional perspective on reinforcement learning.
\newblock In \emph{Proceedings of the 34th International Conference on Machine
  Learning}, ICML, pages 449--458. PMLR, 2017.

\bibitem[Lillicrap et~al.(2016)Lillicrap, Hunt, Pritzel, Heess, Erez, Tassa,
  Silver, and Wierstra]{lillicrap2016ddpg}
Timothy~P. Lillicrap, Jonathan~J. Hunt, Alexander Pritzel, Nicolas Heess, Tom
  Erez, Yuval Tassa, David Silver, and Daan Wierstra.
\newblock Continuous control with deep reinforcement learning.
\newblock In \emph{International Conference on Learning Representations
  ({ICLR})}, 2016.

\bibitem[Fujimoto et~al.(2018)Fujimoto, van Hoof, and Meger]{fujimoto2018td3}
Scott Fujimoto, Herke van Hoof, and David Meger.
\newblock Addressing function approximation error in actor-critic methods.
\newblock In \emph{International Conference on Machine Learning ({ICML})},
  2018.

\bibitem[van~der Ploeg and Venables(2011)]{vanderploeg2011}
Frederick van~der Ploeg and Anthony~J. Venables.
\newblock Harnessing windfall revenues: {Optimal} policies for resource-rich
  developing economies.
\newblock \emph{Economic Journal}, 121\penalty0 (551):\penalty0 1--30, 2011.
\newblock \doi{10.1111/j.1468-0297.2010.02411.x}.

\end{thebibliography}

\end{document}